\documentclass{emulateapj} %[12pt,preprint]{aastex}

\usepackage {epsf}
\usepackage{epstopdf}
\usepackage {lscape, graphicx}
\usepackage {amssymb,amsmath}

\gdef\1054{MS\,1054--03}

\gdef\KMOS3D{\rm KMOS$^{\rm 3D}$}

\def\farcs{\hbox{$.\!\!^{\prime\prime}$}}
\def\simgeq{{\raise.0ex\hbox{$\mathchar"013E$}\mkern-14mu\lower1.2ex\hbox{$\mathchar"0218$}}}

\begin {document}

\title {\KMOS3D: Dynamical Constraints on the Mass Budget in Early Star-forming Disks$^*$} %\thanks{$^*$ Based on observations collected at the European Organisation for Astronomical Research in the Southern Hemisphere under ESO programmes 092.A-0091, 093.A-0079, 094.A-0217, 095.A-0047, and 096.A-0025.}

\author{Stijn Wuyts\altaffilmark{1},
Natascha M. F\"{o}rster Schreiber\altaffilmark{2},
Emily Wisnioski\altaffilmark{2},
Reinhard Genzel\altaffilmark{2,3,4},
Andreas Burkert\altaffilmark{5},
Kaushala Bandara\altaffilmark{2},
Alessandra Beifiori\altaffilmark{2,5},
Sirio Belli\altaffilmark{2},
Ralf Bender\altaffilmark{2,5},
Gabriel B. Brammer\altaffilmark{6},
Jeffrey Chan\altaffilmark{2},
Ric Davies\altaffilmark{2},
Matteo Fossati\altaffilmark{5,2},
Audrey Galametz\altaffilmark{2},
Sandesh K. Kulkarni\altaffilmark{2},
Philipp Lang\altaffilmark{2},
Dieter Lutz\altaffilmark{2},
J. Trevor Mendel\altaffilmark{2},
Ivelina G. Momcheva\altaffilmark{6},
Thorsten Naab\altaffilmark{7},
Erica J. Nelson\altaffilmark{8},
Roberto P. Saglia\altaffilmark{2,5},
Stella Seitz\altaffilmark{5},
Linda J. Tacconi\altaffilmark{2},
Ken-ichi Tadaki\altaffilmark{2},
Hannah \"{U}bler\altaffilmark{2},
Pieter G. van Dokkum\altaffilmark{8},
David J. Wilman\altaffilmark{5,2},
Eva Wuyts\altaffilmark{2}} %\ \altaffilmark{*}}

\altaffiltext{1}{Department of Physics, University of Bath, Claverton Down, Bath, BA2 7AY, UK}
\altaffiltext{2}{Max-Planck-Institut f\"{u}r extraterrestrische Physik, Postfach 1312, Giessenbachstr., D-85741 Garching, Germany}
\altaffiltext{3}{Department of Physics, Le Conte Hall, University of California, Berkeley, CA 94720, USA}
\altaffiltext{4}{Department of Astronomy, Hearst Field Annex, University of California, Berkeley, CA 94720, USA}
\altaffiltext{5}{Universit\"{a}ts-Sternwarte M\"{u}nchen, Scheinerstr. 1, M\"{u}nchen, D-81679, Germany}
\altaffiltext{6}{Space Telescope Science Institute, 3700 San Martin Drive, Baltimore, MD 21218, USA}
\altaffiltext{7}{Max-Planck-Institut f\"{u}r Astrophysik, Karl Schwarzschildstr. 1, D-85748 Garching, Germany}
\altaffiltext{8}{Astronomy Department, Yale University, New Haven, CT 06511, USA}

\altaffiltext{*}{Based on observations collected at the European Organisation for Astronomical Research in the Southern Hemisphere under ESO programmes 092.A-0091, 093.A-0079, 094.A-0217, 095.A-0047, and 096.A-0025.}

\begin{abstract}
We exploit deep integral-field spectroscopic observations with KMOS/VLT of 240 star-forming disks at $0.6 < z < 2.6$ to dynamically constrain their mass budget.  Our sample consists of massive ($\gtrsim 10^{9.8}\ M_{\sun}$) galaxies with sizes $R_e \gtrsim 2$ kpc.  By contrasting the observed velocity and dispersion profiles to dynamical models, we find that on average the stellar content contributes $32^{+8}_{-7}\%$ of the total dynamical mass, with a significant spread among galaxies (68th percentile range $f_{\rm star} \sim 18 - 62\%$).  Including molecular gas as inferred from CO- and dust-based scaling relations, the estimated baryonic mass adds up to $56^{+17}_{-12}\%$ of total for the typical galaxy in our sample, reaching $\sim 90\%$ at $z > 2$.  We conclude that baryons make up most of the mass within the disk regions of high-redshift star-forming disk galaxies, with typical disks at $z > 2$ being strongly baryon-dominated within $R_e$.  Substantial object-to-object variations in both stellar and baryonic mass fractions are observed among the galaxies in our sample, larger than what can be accounted for by the formal uncertainties in their respective measurements.  In both cases, the mass fractions correlate most strongly with measures of surface density.  High $\Sigma_{\rm star}$ galaxies feature stellar mass fractions closer to unity, and systems with high inferred gas or baryonic surface densities leave less room for additional mass components other than stars and molecular gas.  Our findings can be interpreted as more extended disks probing further (and more compact disks probing less far) into the dark matter halos that host them.
\end{abstract}


\section {Introduction}
\label{intro.sec}

Measurements of mass and mass growth with cosmic time are key ingredients to any theory of galaxy formation and evolution.  In the local universe, decades of intensive studies have culminated in a census of the stellar, molecular gas and atomic gas content of galaxies (e.g., Cole et al. 2001;  Keres et al. 2003; Martin et al. 2010), as well as their respective spatial distributions on subgalactic scales (e.g., Zibetti et al. 2009; Leroy et al. 2009; Walter et al. 2008).  Likewise, constraints on the total dynamical mass budget of galaxies from kinematic and/or lensing studies have come a long way since the pioneering work on rotation curves by Rubin et al. (1978) and Bosma (1978), placing constraints on central dark matter fractions and/or stellar mass-to-light ratios of both spiral and elliptical galaxies (see, e.g., Courteau \& Dutton 2015 and references therein; notably Bershady et al. 2011; Martinsson 2013a,b; Barnab\`{e} et al. 2011, 2012; Brewer et al. 2012; Thomas et al. 2011; Cappellari et al. 2012, 2013, 2016).  Indications from the above studies are that nearby spiral disks are baryon-dominated at the center and dark-matter dominated in their outskirts, with the baryonic mass fraction at a given radius being larger for more massive systems.  The stellar initial mass function (IMF) may vary from a Chabrier (2003) IMF in spiral disks to a more bottom-heavy, Salpeter (1955) IMF in the central regions of massive ellipticals (see also van Dokkum \& Conroy 2010).

In comparison, efforts to establish a census of the mass budget in (and the spatial distribution of its respective components within) galaxies at higher lookback times are still in their infancy.  Among the various baryonic constituents, estimates of the stellar mass content have probably matured most.  Extensive multi-wavelength imaging data sets such as provided by the UltraVISTA (McCracken et al. 2012) and CANDELS (Grogin et al. 2011; Koekemoer et al. 2011) surveys have yielded a mass-complete census of galaxy-integrated stellar masses out to the peak of cosmic star formation more than 10 billion years ago (e.g., Ilbert et al. 2013; Muzzin et al. 2013), and even shed light on the spatial distribution of stellar mass within those early galaxies (Wuyts et al. 2012; Lang et al. 2014).  Nevertheless, the same caveats as present locally, regarding systematic uncertainties in the derived stellar masses, apply to all lookback studies.  These include, but are not limited to, the adopted IMF, assumptions regarding star formation histories and dust attenuation, and calibrations of the input stellar population synthesis (SPS) models. 

Progress in understanding the molecular gas content of distant galaxies has come more recently, with the focus of the sub-mm community shifting gradually from rare, exceptionally luminous infrared galaxies to samples that are inherently fainter but more representative of the general underlying population.  While direct measurements of the molecular gas mass function remain a challenge, impressive leaps forward have been made in establishing cold gas scaling relations based on CO and dust observations that can be used to `populate' the complete underlying galaxy population with molecular gas masses (Genzel et al. 2015; see also Tacconi et al. 2010, 2013; Daddi et al. 2010a,b; Carilli \& Walter 2013).  These observations showed that molecular gas is an increasingly important contributor to the total mass budget within galaxies at higher lookback times (see also Berta et al. 2013), with gas mass fractions of $\sim 0.33$ at $z \sim 1.2$ and $\sim 0.47$ at $z \sim 2.2$ (e.g., Tacconi et al. 2013).  Here, too, certain caveats apply, such as the calibration of the CO-to-$H_2$ conversion factor (e.g., Wolfire et al. 2010; Krumholz et al. 2011; Genzel et al. 2012; Magnelli et al. 2012; Bolatto et al. 2013) and the excitation correction (e.g., Narayanan \& Krumholz 2014) on the CO side, and the dust-to-gas ratio for inferences based on the far-infrared continuum (e.g., Leroy et al. 2011; R\'{e}my-Ruyer et al. 2014; Groves et al. 2015).  On a spatially resolved level, the first attempts to map the molecular gas mass distribution within distant galaxies have recently been presented (Tacconi et al. 2013; Genzel et al. 2013; Freundlich et al. 2013), with more to come from ALMA.

While due to the high pressure of the interstellar medium (ISM) atomic gas is believed to be a relatively minor ingredient within the visible extent of the high-surface density galaxies observed at redshifts $z \sim 1$ to 2, a direct confirmation of this expectation has to await the Square Kilometer Array (and out to $z \sim 1$ its pathfinders MEERKAT and ASKAP). 

In light of the potential systematics involved in estimating both the stellar and gaseous components (and their spatial distribution), independent dynamical measurements provide a welcome 'reality check'.  Conversely, given sufficient precision (or trust) in the baryonic measurements, kinematic constraints on the enclosed mass may reveal the presence of hidden mass such as dark matter, and provide a test of the stellar initial mass function.  At the minimum, baryonic masses should not grossly exceed the total amount of mass inferred dynamically to be present within a given radius.

For quiescent galaxies, such tests have recently been carried out out to $z \sim 2$, suggesting stellar mass fractions of order 50\% (modulo IMF uncertainties), with a substantial object-to-object scatter (van de Sande et al. 2013; Bezanson et al. 2013).  Here, the typical approach is to assume spherical symmetry, and adopt a geometric factor linking the direct observables (galaxy size and velocity dispersion) to the dynamical mass ($M_{dyn}$).  Possible rotational components (Newman et al. 2015) and velocity dispersion anisotropies (van der Marel 1991) may in reality complicate the interpretation of quiescent galaxy kinematics, but are by lack of constraints rarely incorporated in high-redshift studies.

An additional factor of consideration in the case of star-forming galaxies, is that their mass budget has a significant contribution by a gaseous component, negligible in quiescent systems.   Moreover, star-forming galaxies (SFGs) are not spherical systems dominated by random motions, but generally show patterns of ordered disk rotation, albeit at high redshift with a significant dispersion and hence non-negligible pressure support component (F\"{o}rster Schreiber et al. 2009; Wisnioski et al. 2015), making the translation between the observed line-of-sight velocity and the enclosed dynamical mass sensitive to inclination.  Precise determinations of the inclination can furthermore be complicated by the often clumpy morphological appearance of high-redshift galaxies, even when they exhibit ordered disk rotation (F\"{o}rster Schreiber et al. 2011).

As we will argue in this paper, the degeneracy between mass and inclination impacting observed radial velocities can in part be addressed by analyzing the kinematic properties of large samples of homogeneously selected distant galaxies.  The advent of KMOS, a new multi-object near-infrared integral-field spectrograph on the VLT, makes such an approach uniquely possible (see also Stott et al. 2016).  Multi-object slit-based spectroscopy with MOSFIRE provides an alternative means to investigating the mass budget in distant galaxies (Price et al. 2016), and complementary VLT/MUSE observations have shed light on the lower mass counterparts (Contini et al. 2016).  Here, we exploit deep KMOS imaging spectroscopy obtained during the first 5 semesters of the \KMOS3D\ program (PI N. M. F\"{o}rster Schreiber; Wisnioski et al. 2015) to contrast the observed H$\alpha$ kinematics of 240 star-forming disks at $0.6 < z < 2.6$ to dynamical models based on their stellar mass and gas content, informed by high-resolution HST CANDELS observations of their structure.  

The paper is structured as follows.  In Section\ \ref{obs_sample.sec}, we briefly describe the observations and lay out the selection of our sample.  In Section\ \ref{methodology.sec}, we give an overview of the methodologies used to construct velocity and dispersion profiles, model the dynamics, and estimate the baryonic mass content.  We present distributions of stellar and baryonic mass fractions, as well as relations between dynamical and stellar/baryonic masses in Section\ \ref{massbudget.sec}.  Section\ \ref{incdistr.sec} addresses whether uncertainties and potential biases in inclination can account for the inferred missing mass, and Section\ \ref{dependence.sec} investigates the relation between mass fractions and other galaxy properties such as redshift and surface density.  We discuss the presence of objects with baryonic masses exceeding the dynamical constraints in Section\ \ref{discussion_fbargt1.sec}, and view our findings in the light of expectations from the cosmological hydrodynamical simulation Illustris in Section\ \ref{discussion_illustris.sec}.  Finally, we summarize our findings on the mass budget in early star-forming disks in Section\ \ref{summary.sec}.

Throughout this paper, we assume a Chabrier (2003) initial mass function (IMF) and
adopt the following cosmological parameters: $(\Omega _M, \Omega_{\Lambda}, h) = (0.3, 0.7, 0.7)$.

%%%%%
% FIG 1
%%%%%
\begin {figure*}[t]
\epsscale{1.0} %{1.1}
\plotone{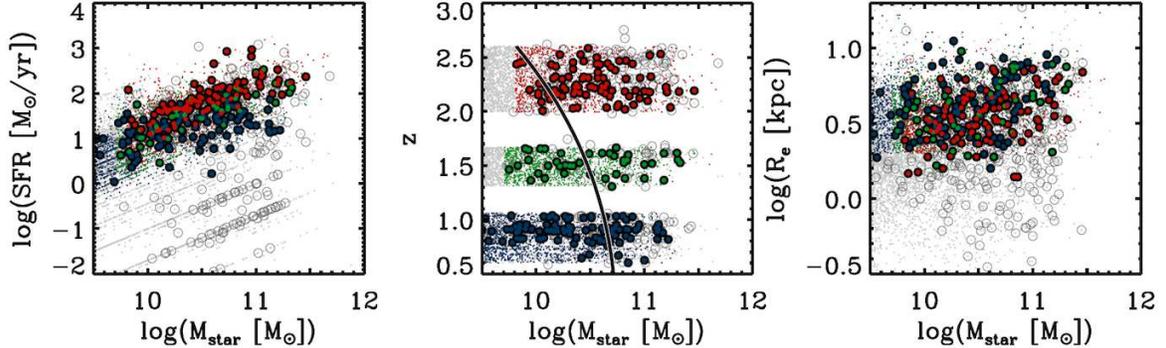}
\epsscale{1.0}
\caption{Our kinematic sample of SFGs at $z \sim 0.9$ (blue filled circles), $z \sim 1.5$ (green filled circles) and $z \sim 2.3$ (red filled circles) in diagrams of SFR, redshift, and size as a function of stellar mass.  For reference, open circles mark the remaining \KMOS3D\ galaxies with H$\alpha$ detections observed between October 2013 and January 2016, and dots represent the underlying galaxy population in the same mass and redshift range.  Our galaxies span the entire 'main sequence' of star formation.  The solid line in the middle panel marks an evolving mass limit corresponding to a fixed cumulative comoving number density of the underlying population of $10^{-3}$ Mpc$^{-3}$ (see Section\ \ref{redshift.sec}).
\label{sample.fig}}
\vspace{-0.5cm}
\end{figure*}

\section{Observations and Sample}
\label{obs_sample.sec}

\subsection{\KMOS3D}
\label{KMOS3D_obs.sec}

The central ingredient of our analysis is the kinematic information, using H$\alpha$ as a tracer, provided by the \KMOS3D\ survey (PI N. M. F\"{o}rster Schreiber).  For an in depth description of the survey strategy, data quality and handling, we refer the reader to Wisnioski et al. (2015).  Here, we briefly summarize some of its key features.  \KMOS3D\ targets spanning a wide dynamic range in mass, star formation rate (SFR) and color were selected from the three CANDELS/3D-HST fields within reach from the VLT: GOODS-South, COSMOS, and UDS.  High-quality grism redshifts by 3D-HST (Brammer et al. 2012; Momcheva et al. 2016) guarantee the H$\alpha$ emission line to be free of OH sky line contamination\footnote{We emphasize that we do not restrict ourselves to sources with emission lines detected in the grism data, but also use grism redshifts based on continuum information.  The 700 - 1000 km/s precision of the grism redshifts is adequate for the OH avoidance criteria.}.  We target $z \sim 0.9$ galaxies using the $YJ$ grating, $z \sim 1.5$ galaxies in $H$, and $z \sim 2.3$ galaxies in $K$. The observations are deep, ranging from 3 to 20 hours on source, with median exposure times of 4.7 hours in $YJ$, 7.8 hours in $H$, and 8.3 hours in $K$.  Typical seeing conditions varied from $0\farcs4$ to $0\farcs8$ in the near infrared, with a median for the data set explored here of $0\farcs5$, as monitored real time by PSF stars positioned on three of the 24 IFU arms.  The data reduction was carried out using the standard software package for KMOS: SPARK (Davies et al. 2013), complemented with in-house custom IDL codes.  Each KMOS IFU has a $2\farcs8 \times 2\farcs8$ footprint.  Dithered observations and minor offsets in pointing between different runs or observing blocks yield typical field of views of $3\farcs4$ to $4''$ on a side for each target.  Any noise enhancements in the lesser exposed outer pixels are propagated to the kinematic extraction.

\subsection{Ancillary Data}
\label{ancillary.sec}

In addition to the KMOS$^{\rm 3D}$ observations, our analysis fundamentally relies on the wealth of multi-wavelength data already available in the CANDELS/3D-HST fields.  We make use of the multi-wavelength optical-to-8$\mu$m photometric catalogs by the 3D-HST team (Skelton et al. 2014), extended with the {\it Spitzer}/MIPS and {\it Herschel}/PACS photometry at longer wavelengths provided by Whitaker et al. (2014) and PEP + GOODS-Herschel (Lutz et al. 2011; Magnelli et al. 2013).  Stellar masses, SFRs and other galaxy properties based thereupon were derived following identical procedures as outlined by Wuyts et al. (2011b).  A final key ingredient to our analysis is the WFC3 data obtained by the CANDELS multi-cycle program (Grogin et al. 2011; Koekemoer et al. 2011), providing a high-resolution view on the rest-frame optical structure of our galaxies.

\subsection{Sample definition} 
\label{sample.sec}

%%%%%
% FIG 2
%%%%%
\begin {figure}[t]
\epsscale{1.14}
\plotone{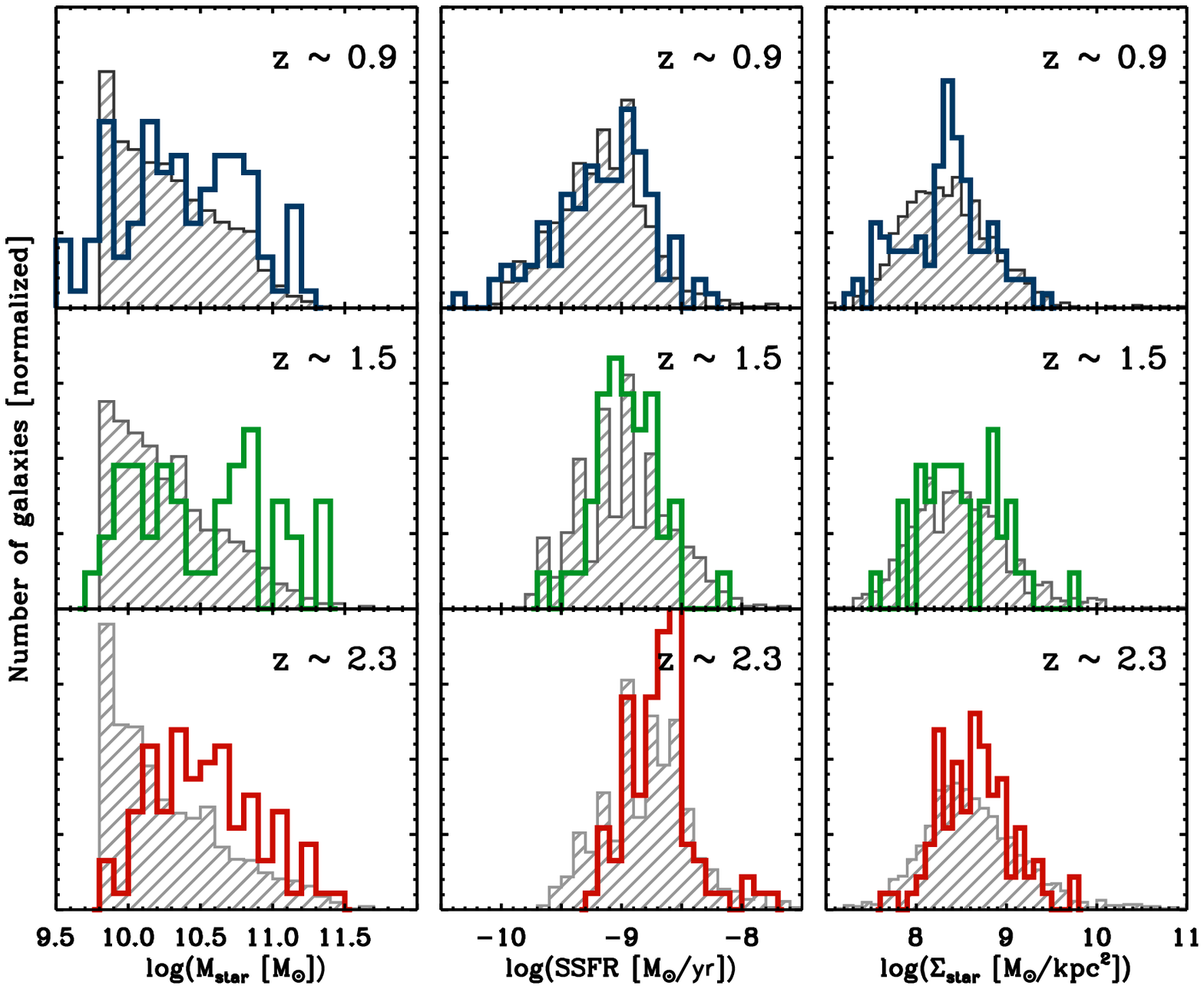}
\plotone{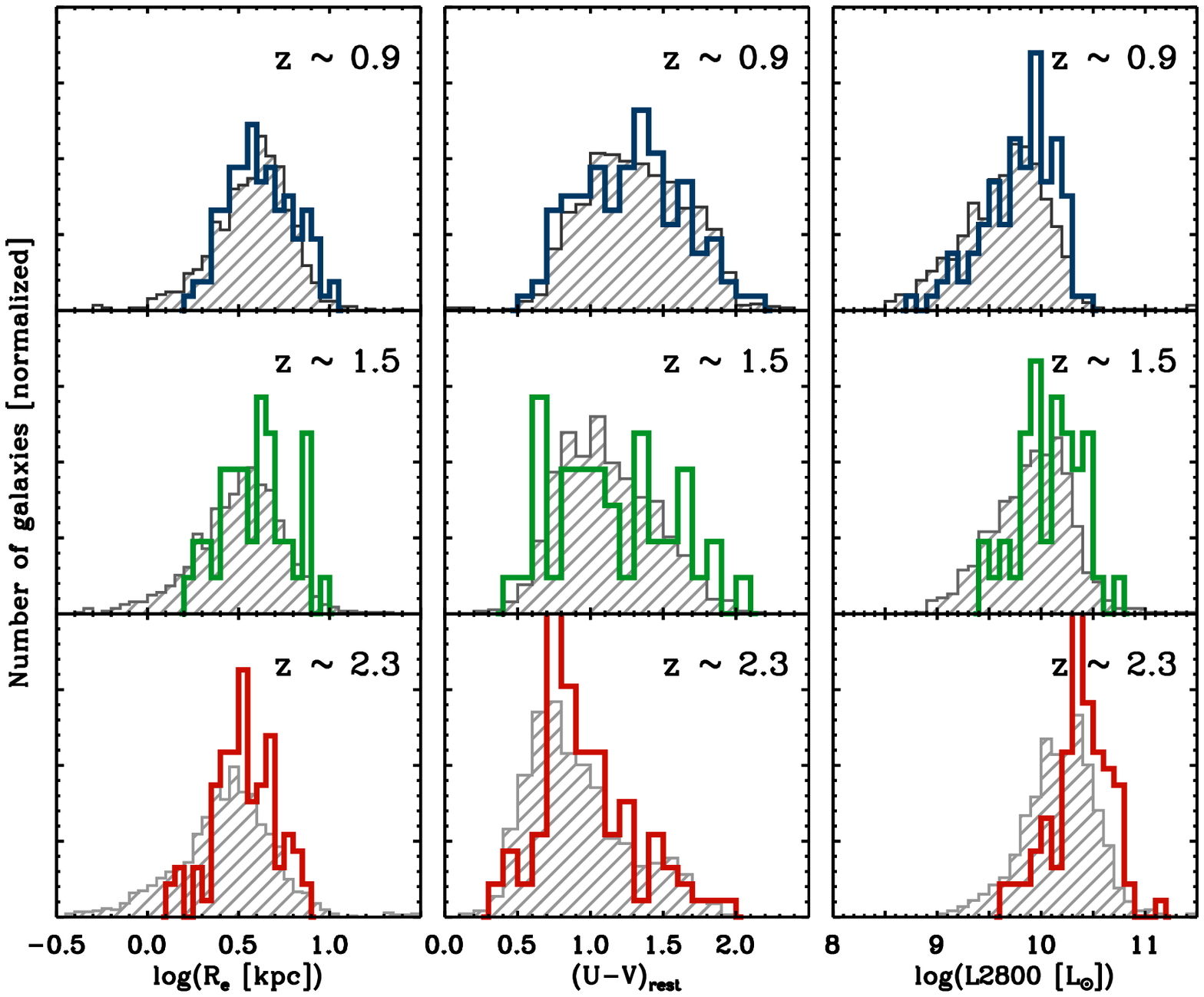}
\epsscale{1.0}
\caption{Normalized distribution of galaxy properties for our kinematic sample, contrasted to a mass-complete parent population above $\log(M_{star}) > 9.8$.  {\it Top:} Stellar mass, specific star formation rate, stellar surface density.  {\it Bottom:} $H$-band effective radius, rest-optical color, rest-UV luminosity.  The kinematic sample spans a similar range in properties as featured in the underlying population, with generally subtle differences in the distributions detailed in the text.
\label{representative.fig}}
\vspace{-0.5cm}
\end{figure}

The total sample of galaxies for which H$\alpha$ emission was detected during our 10/2013 - 01/2016 observations counts 407 targets (a success rate of $>75\%$).  At least 70\% of the KMOS$^{\rm 3D}$ sample satisfies disk classification criteria as listed by Wisnioski et al. (2015): most notably a monotonic velocity gradient and centrally peaked velocity dispersion map.  For this study, we conservatively selected a sample of 240 galaxies with velocity gradients along a clearly identified kinematic major axis, and sufficient signal-to-noise and spatial extent in H$\alpha$ to merit a detailed kinematic analysis including fitting of disk dynamical models.  106 of these reside in the redshift range $0.6 < z < 1.1$.  The $1.3 < z < 1.7$ bin counts 42 galaxies (as $H$ band observations started later in our program). The remaining 92 sources span the range $2.0 < z < 2.6$.  The signal-to-noise ratio of the H$\alpha$ line emission ranges from $S/N \sim 10 - 100$ in the galaxy centers to $S/N \sim 5$ in the outermost radial bins, which on average reach 2.5$\times$ the effective $H$-band radius.

Figure\ \ref{sample.fig} showcases some of their key properties.  The galaxies we analyze sample the SFR - mass `main sequence' relation (Noeske et al. 2007; Daddi et al. 2007; Elbaz et al. 2007) at their respective redshifts ($SSFR \gtrsim 0.7 / t_{\rm Hubble}$), and are broadly confined to stellar masses above $\sim 10^{9.8}\ M_{\sun}$ and sizes (major axis effective radii in the $H$ band) larger than $\sim 2$ kpc.  Within this region of parameter space, the KMOS$^{\rm 3D}$ success rate in obtaining H$\alpha$ detections is 92\%.

A further 89 H$\alpha$-detected \KMOS3D\ galaxies fall in the same region of SFR - mass and size - mass parameter space, but were excluded from our sample.\footnote{Also excluded from the original sample of 407 galaxies with H$\alpha$ detections were 78 galaxies that cover different regions of parameter space, at lower mass (13/78), lower SSFR (22/78) and/or smaller size (61/78; 44 of which have $\log(M_{star}) > 9.8$ and $\log(SSFR) > 0.7/t_{Hubble}$).  Their characteristics will be the focus of future studies, as number statistics for these populations build up and deeper data is obtained.}  In roughly equal numbers, the reason for these conservative cuts were insufficient signal-to-noise, the absence of a clean rotational pattern with unambiguous kinematic axis, and offsets between kinematic and morphological position angles exceeding 40$^{\circ}$.  The latter often coincide with orientations that are too face-on to be constraining in terms of disk dynamical modeling (see also Wisnioski et al. 2015).  The axial ratio distribution of excluded objects in the same region of SFR-size-mass parameter space as our core sample of 240 galaxies spans a large range from 0.2 to 1.0.  It is slightly skewed toward rounder shapes for the aforementioned reason.

Six objects were weeded out on the basis of a clear merger morphology for which no meaningful axial ratio (and hence inclination) could be measured (following guidelines as outlined by Kartaltepe et al. 2015), and in which case the framework of disk rotation probing the potential well depth may in any case not be appropriate.  We note that a fraction of the galaxies in our sample may be undergoing minor mergers, which can be hard to distinguish from clumps formed in situ due to instabilities in gas-rich disks (see, e.g., Mandelker et al. 2014).  Mock observations of hydrodynamical simulations illustrate how during certain stages and under certain viewing angles the orbital motions of such minor mergers may mimic signatures of disk rotation (Simons et al. in prep).  Depending on how aggressively or conservatively one selects, we estimate that 6 - 15\% of our sample could tentatively be identified as a minor merger.  We verified that this subpopulation does not occupy a unique corner of parameter space nor deviates from the general trends described in this paper, and excluding them would consequently not alter our conclusions.

To evaluate how representative our sample is for the overall population of massive star-forming galaxies, we extract a mass-complete sample of SFGs ($\log(M_{star}) > 9.8$ \& $SSFR > 0.7 / t_{\rm Hubble}(z)$) from the 3D-HST/CANDELS data set (Skelton et al. 2014; Momcheva et al. 2016).  In Figure\ \ref{representative.fig}, we show their distribution in stellar mass, specific star formation rate, stellar surface density (a key parameter discussed in Section\ \ref{surfdens.sec}), size, rest-optical color, and rest-frame UV (2800\AA) luminosity, alongside that of the kinematic sample.  Whereas there is a large overlap between the distributions, and in many cases even a tight match, the kinematic sample is not drawn randomly from a mass-complete population of SFGs.  The kinematic sample features a flatter mass distribution than an exponentially declining Schechter function.  This is a deliberate choice in the survey design, where, particularly in the early runs, an emphasis was placed on the more massive galaxies.  Furthermore, galaxies with bright rest-UV luminosities are slightly overrepresented compared to the underlying population.  This reflects the high demand on data quality and H$\alpha$ signal-to-noise ratio (the latter not being available for full underlying 3D-HST sample, but correlating with $L_{2800}$).  A Kolmogorov-Smirnov test confirms that, in each redshift bin, the probability that the kinematic sample and the mass-complete 3D-HST sample share the same parent stellar mass and $L_{2800}$ distribution is less than 2\%.  A minor deficit of low SSFR, and very small systems is only statistically significant in the highest redshift bin.  In Section\ \ref{redshift.sec}, after showing the results for the kinematic sample, we apply weights to galaxies in the KMOS$^{\rm 3D}$ sample in order to address to which degree median mass fractions and their redshift evolution may differ for a mass-complete population of SFGs.  Any changes found are at the level of 0.1 dex or smaller.

\section {Methodology}
\label{methodology.sec}

The approach we take in this paper is not to fit a full-fledged dynamical model to the observed kinematics, leaving free a large number of parameters (e.g., total mass, inclination, spatial mass and light distribution) and possibly degeneracies between them.  Instead, we construct dynamical models based on the inclination and size of the galaxies, known from rest-optical HST imagery.  We fit those to the observed kinematics, leaving just two parameters free: the (dynamical) mass $M_{dyn}$, and an intrinsic velocity dispersion $\sigma_0$ representing the floor of the radial dispersion profile and associated degree of pressure support (see\ \ref{coupling.sec}).  In the following, we first outline the kinematic extraction (Section\ \ref{velprof.sec}), then describe in more detail how we computed the dynamical masses (Section\ \ref{dysmal.sec}), and briefly summarize the methodology used to infer the stellar and gas masses that enter the assessment of the baryonic mass fractions\footnote{Note that throughout this paper, what we consider is the mass fraction {\it within} the disk.  For a detailed analysis of the mass fractions with respect to the total, larger scale dark matter halo in which the galaxies are embedded, see Burkert et al. (2016).  We also assume the baryonic mass to equal the sum of stellar mass and molecular gas mass (see also Section\ \ref{discussion_fbargt1.sec}).} (Sections\ \ref{Mstar.sec} and\ \ref{Mgas.sec}).

\subsection{Velocity and Dispersion Profiles along the Kinematic Major Axis}
\label{velprof.sec}

\begin {figure*}[htbp]
\plotone{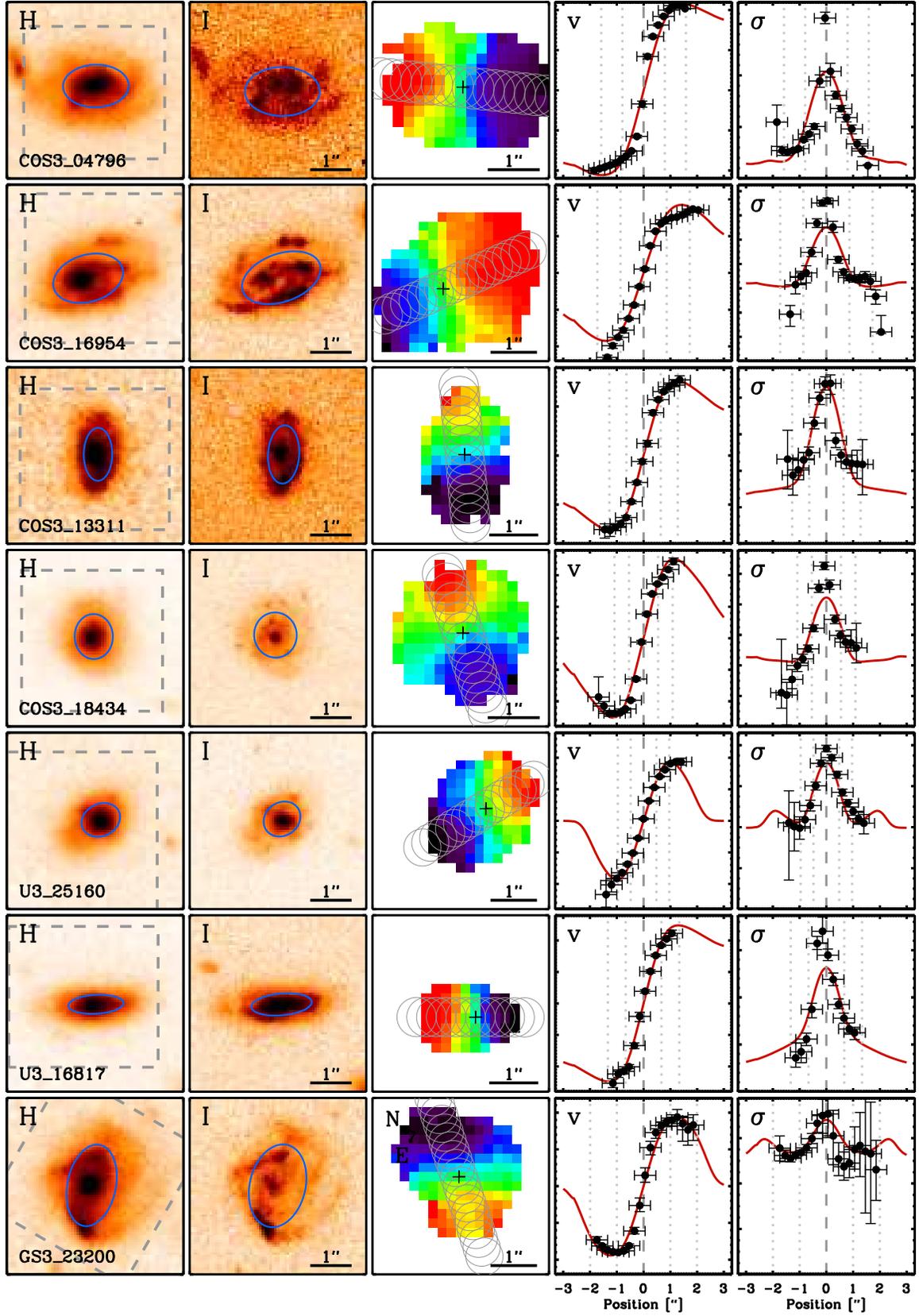}
\caption{Case examples of high-redshift galaxies showing ordered disk rotation.  From left to right: surface brightness distribution in the WFC3 $H$ and ACS $I$ band, with blue ellipses indicating the GALFIT effective radius and gray dashed lines marking the field of view of KMOS observations; $H\alpha$ velocity field with circles marking the extracted pseudo-slit; the observed and modeled 1D velocity and velocity dispersion profile along the major axis.
\label{gallery.fig}}
\end{figure*}

We construct radial velocity and dispersion profiles by running LINEFIT on a series of spectra extracted from the 3D data cube within circular apertures of $0\farcs8$ in diameter that were placed along the kinematic major axis.  The LINEFIT code (Davies et al. 2009; F\"{o}rster Schreiber et al. 2009) fits a line profile to the continuum-subtracted spectral profile, implicitly accounting for the spectral resolution.  The uncertainties are boot-strapped using Monte Carlo techniques.  Examples of extracted rotation and dispersion profiles are presented in Figure\ \ref{gallery.fig}.

While employing a pseudo-slit extraction, we emphasize that the analysis presented in this paper benefits critically from the integral-field nature of our data set, enabling a more reliable kinematic classification as well as a more reliable determination of the kinematic center and major axis position angle, which is critical in deriving dynamical masses (e.g., Swaters et al. 2003; F\"{o}rster Schreiber et al. 2009; Wisnioski et al. 2015).  For the purpose of our analysis, and given the combination of angular resolution and size of our galaxies, the major axis information provides the strongest constraints on the simple disk models we use.  In contrast, the off-axis information is relatively insensitive to constrain the velocity gradient, and depends mainly on inclination (see, e.g., Figure 1 of van der Kruit \& Allen 1978 and Figure 9 of Glazebrook 2013).  Since we derive the inclination from independent constraints, namely HST axial ratios, we chose to follow Genzel et al. (2014a) and Wisnioski et al. (2015) in adopting the pseudo-slit approach.  We note that in the absence of independent inclination constraints a full fitting in 3D space may be preferred, as suggested by the analyses of Bouch\'e et al. (2015) and Contini et al. (2016).

\subsection{Dynamical Mass Modeling}
\label{dysmal.sec}

We carry out a forward modeling procedure, fitting the velocity and velocity dispersion profiles extracted along the major kinematic axis simultaneously in observed space, through standard least squares minimization.  In each iteration, the two free parameters $M_{dyn}$ and $\sigma_0$ are varied and an expected observed rotation curve and dispersion profile is computed, using the same circular aperture extraction technique as described in Section\ \ref{velprof.sec}, and accounting for beam smearing with a PSF appropriate for the observations of the galaxy in question.  To this end, we use an updated version of DYSMAL (Davies et al. 2011; see also Cresci et al. 2009), a code that quantifies the impact of spectral and spatial beam smearing on a rotating disk given a specified intrinsic mass model, a spatial distribution of the light emitter tracing the gravitational potential, and an inclination with respect to the observer.

%%%%%
% FIG 3a
%%%%%

%%%%%
% FIG 3b
%%%%%
\addtocounter{figure}{-1}
\begin {figure*}[htbp]
\plotone{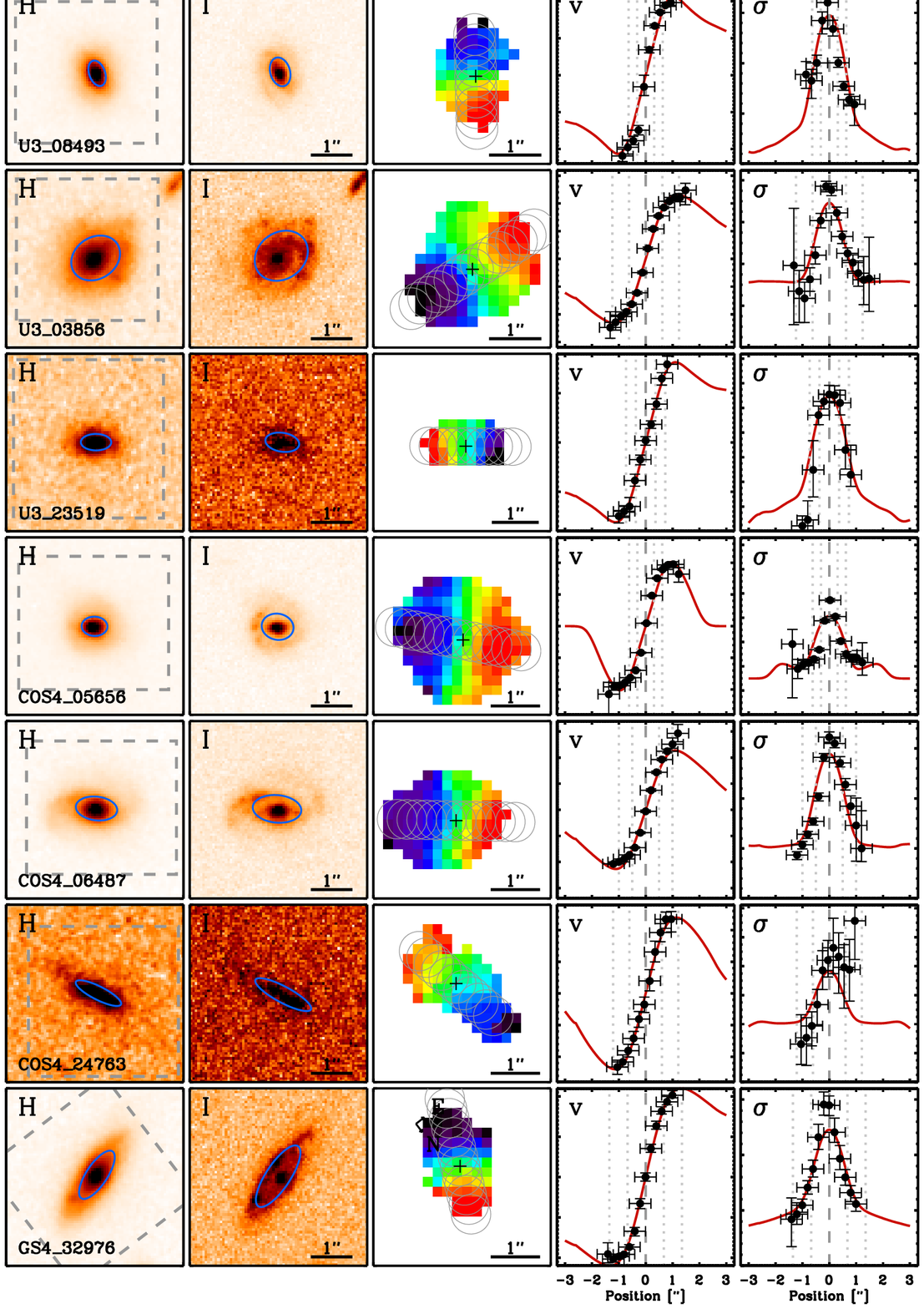}
\caption{Case examples, continued.
}
\end{figure*}

%%%%%
% FIG 3c
%%%%%
\addtocounter{figure}{-1}
\begin {figure*}[htbp]
\plotone{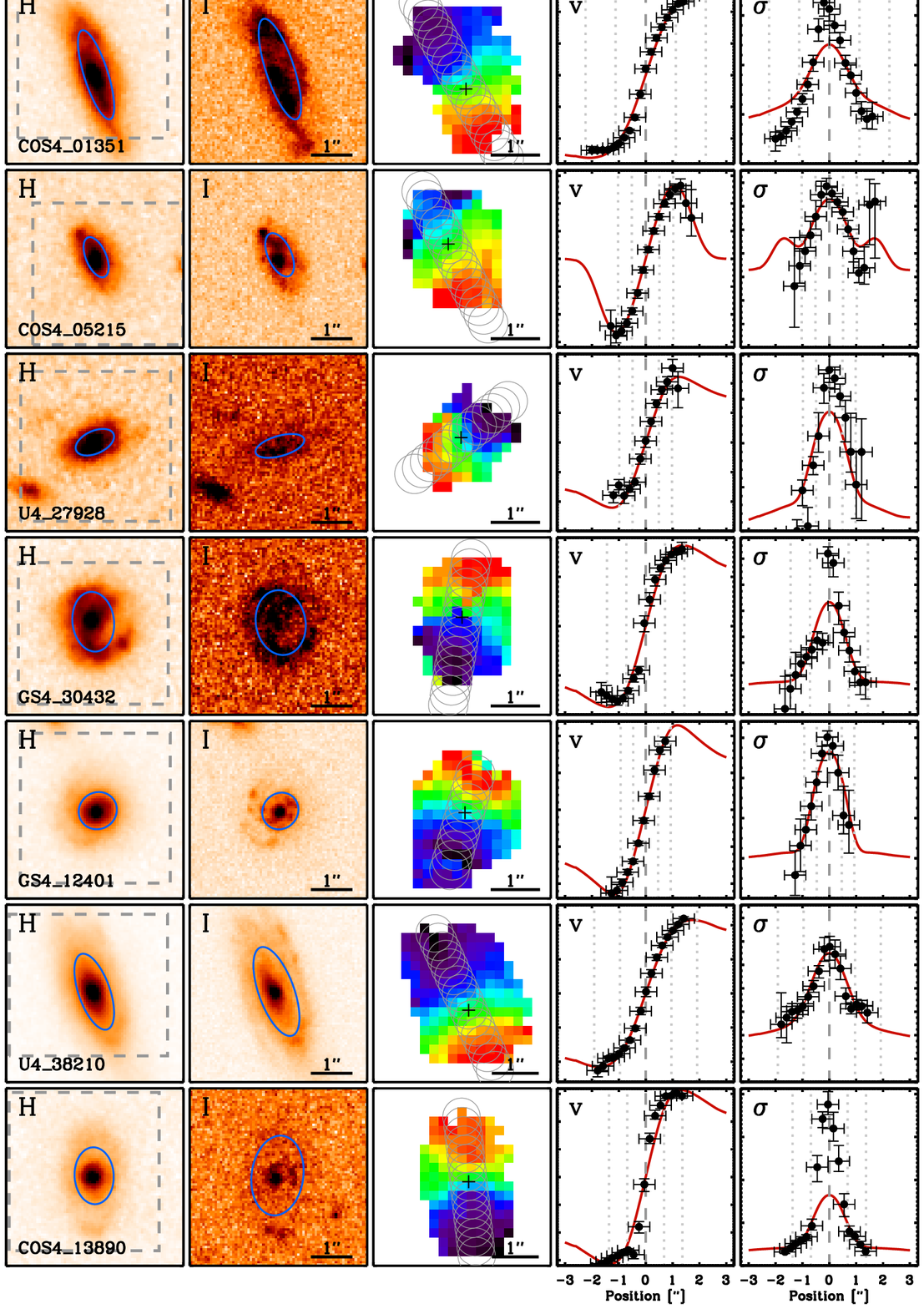}
\caption{Case examples, continued.
}
\end{figure*}

%%%%%
% FIG 3d
%%%%%
\addtocounter{figure}{-1}
\begin {figure*}[htbp]
\plotone{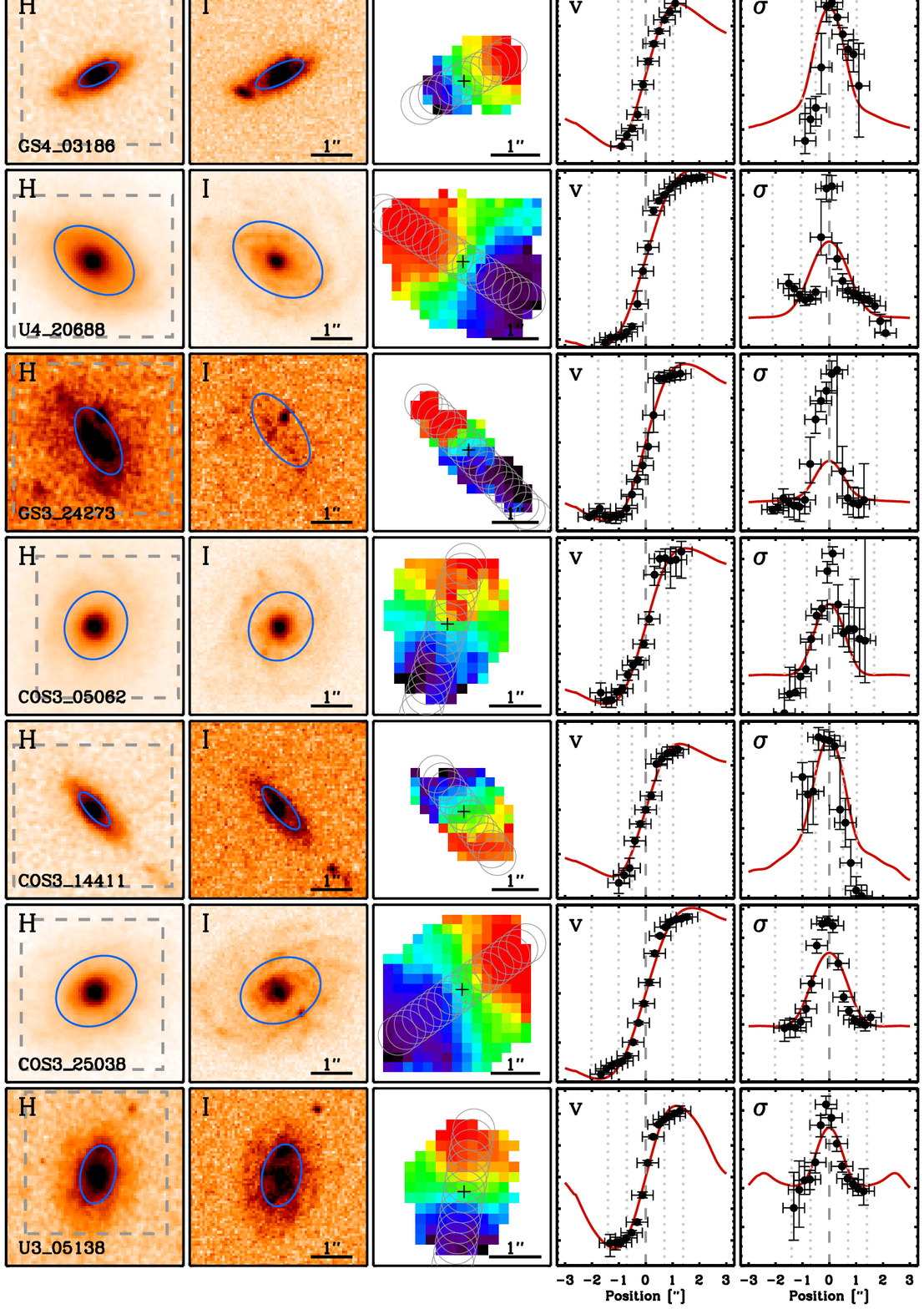}
\caption{Case examples, continued.
}
\end{figure*}

\subsubsection{Mass Distribution}
\label{massdistr.sec}
As spatial distribution of both mass and light we adopt an exponential disk with an effective radius as measured with GALFIT (Peng et al. 2010) on the CANDELS $H$-band image (van der Wel et al. 2012).  The median $H$-band Sersic index of the galaxies in our sample is $n = 1.1$, and the characteristic $H\alpha$ surface brightness distributions of high-redshift SFGs are also well described by exponential disk profiles (Nelson et al. 2013, 2016; F\"{o}rster Schreiber et al. 2016).  We tested that adopting the measured $H$-band Sersic indices (not necessarily a better proxy for the total mass distribution as often half of the baryonic mass is in molecular gas) does not alter our conclusions.  Changes in $M_{dyn}$ for individual objects remain within $\pm 0.05$ dex in 83\% of the cases, and the median mass fraction in any redshift bin changes by no more than 0.01 dex.

In reality, there may be subtle differences between the profile shape and extent of the $H$-band light, $H\alpha$ emission, the stellar mass and molecular gas mass distribution (see, e.g., Wuyts et al. 2012; Nelson et al. 2016; Tacchella et al. 2015; Tadaki et al. in prep), but observational constraints on the combination of these effects on an individual galaxy basis are to date either missing or lacking sufficient robustness to merit a further refinement of our default modeling.  

As a sanity check, we repeated our fitting adopting a more extended mass distribution following the Nelson et al. (2016) scaling $R_{e, H\alpha} = 1.1 R_{e, H} (M/10^{10} M_{\sun})^{0.054}$, based on high-resolution HST grism observations.  As a result, the inferred total dynamical mass increases by $0.08 \pm 0.12$ dex (median and scatter among galaxies in our sample).  However, the mass enclosed within the $H$-band half-light radius $R_{e, H}$ remains robust to $0.01 \pm 0.03$ dex.  We also explored how allowing for the presence of a compact bulge would impact our results.  To this end, we adopted a parameterization in which the mass of a 1 kpc de Vaucouleurs ($n = 4$) component was introduced as an extra free parameter.  In about half of the cases, more than 10\% of the mass is assigned to such a bulge component.  Reassuringly, the quantity of relevance to our present study, the enclosed dynamical mass within $R_{e, H}$, shows only subtle changes: $\Delta \log M_{dyn}(<R_{e,H}) = -0.02 \pm 0.04$ dex.  Finally, we tested the robustness of our results against fitting a superposition of an exponential disk and a NFW halo, leaving $M_{disk}$, $\sigma_0$ and $M_{halo}$ as free parameters.  Here, we fixed the halo concentration to 4, and let the halo's virial radius scale with $M_{halo}$ as in Mo et al. (1998).  Doing so, we again find that, while the total mass of the best-fit model integrated to infinity can be substantially larger than what is obtained in our default modeling, the inferred dynamical mass enclosed within $R_{e, H}$ is robust: $\Delta \log M_{dyn}(<R_{e,H}) = 0.01 \pm 0.02$ dex.

Finally, we tested how deviations from our default assumption, that half of $M_{star}$ and $M_{gas}$ is contained within $R_{e,H}$ would impact the inferred baryonic mass fraction within this radius.  Adopting the Nelson et al. (2016) scaling as a description for how the gaseous extent compares to the $H$-band size, we find median baryonic mass fractions to decrease by -0.02, -0.04, and -0.05 dex at $z \sim 0.9$, 1.5, and 2.3 respectively.  Folding in the presence of stellar $M/L_H$ ratio gradients as derived by Lang et al. (2014; i.e., stars being more centrally concentrated than the $H$-band light) would lead to compensating offsets, by +0.07, +0.05, and +0.04 dex at $z \sim 0.9$, 1.5, and 2.3 respectively.  We conclude that the net effect of the two combined is not expected to significantly affect the results presented in this paper.

\subsubsection{Inclination}
We infer the inclination $i$ from the axial ratio $b/a$ of the WFC3 $H$-band image:

\begin{equation}
\cos{i} = \sqrt{ \frac{(b/a)^2 - thick^2}{1 - thick^2} },
\label{inc.eq}
\end{equation}

where we assumed a ratio of scale height to scale length of $thick = 0.25$.  The latter thickness is consistent with the fall-off at small $b/a$ of the axial ratio distribution constructed for large samples of SFGs at the nominal redshift considered here (van der Wel et al. 2014).  For the 5\% of objects for which the $H$-band axial ratio is marginally smaller than 0.25, we assigned an edge-on inclination.  We tested that inclinations derived from the ACS $I$-band imaging yield similar results, with shifts in the median stellar or baryonic mass fractions limited to $\sim 0.02$ dex.

\subsubsection{Coupling between $v(r)$ and $\sigma(r)$}
\label{coupling.sec}

In computing the intrinsic rotation curve, we account for a finite flattening of the mass distribution following Noordermeer (2008) (i.e., the same $thick = 0.25$ adopted in equation\ \ref{inc.eq}).  We furthermore account for the fact that the shape of the velocity and velocity dispersion profiles are coupled in two ways: through beam smearing and pressure support (see Burkert et al. 2016 for a detailed discussion).  The first is a purely observational effect of finite resolution in which the observed velocity gradient is reduced with respect to the intrinsic one, with beam-smeared velocities giving rise to a central peak in dispersion superposed on a dispersion floor $\sigma_0$.  As illustrated by Burkert et al. (2016), its impact is a steep function of the ratio between beam size and galaxy size.  In the extreme case of an unresolved observation, it implies a velocity gradient is no longer observable and all information on the enclosed dynamical mass is embedded in the velocity dispersion measurement.  

The second effect is intrinsic.  Early star-forming disks are more turbulent and kinematically thicker than nearby spirals (see, e.g., F\"{o}rster Schreiber et al. 2009; Kassin et al. 2012; Wisnioski et al. 2015).  This implies that their dynamical support has a non-negligible pressure component, which has the effect of reducing the rotational velocity $v_{rot}$ compared to the circular velocity $v_{circ}$ of a thin disk in the absence of pressure support, particularly at large radii:

\begin{equation}
v_{rot}^2 = v_{circ}^2 - 2 \sigma_0^2 \left(\frac{r}{R_d}\right),
\end{equation}

where $R_d$ is the disk scale length (Burkert et al. 2010; see also Lang et al., in prep).

Taking into account the $v(r)$ and $\sigma(r)$ constraints simultaneously in a self-consistent manner enhances the robustness of the best-fit dynamical mass within $R_e$, also in cases where a turnover in velocity is not or only marginally detected (see also Appendix A).  Given that we keep the shape of the mass distribution fixed, higher $M_{dyn}$ than those resulting from the fit would lead to velocity gradients that are too steep and, to the degree this is washed out by beam smearing, a central peak in velocity dispersion that is higher than observed.  Robustness against different assumptions on the shape of the mass distribution was addressed in Section\ \ref{massdistr.sec}.

%%%%%
% FIG 4 [v2rat]
%%%%%
\begin {figure}[t]
\epsscale{1.14}
\plotone{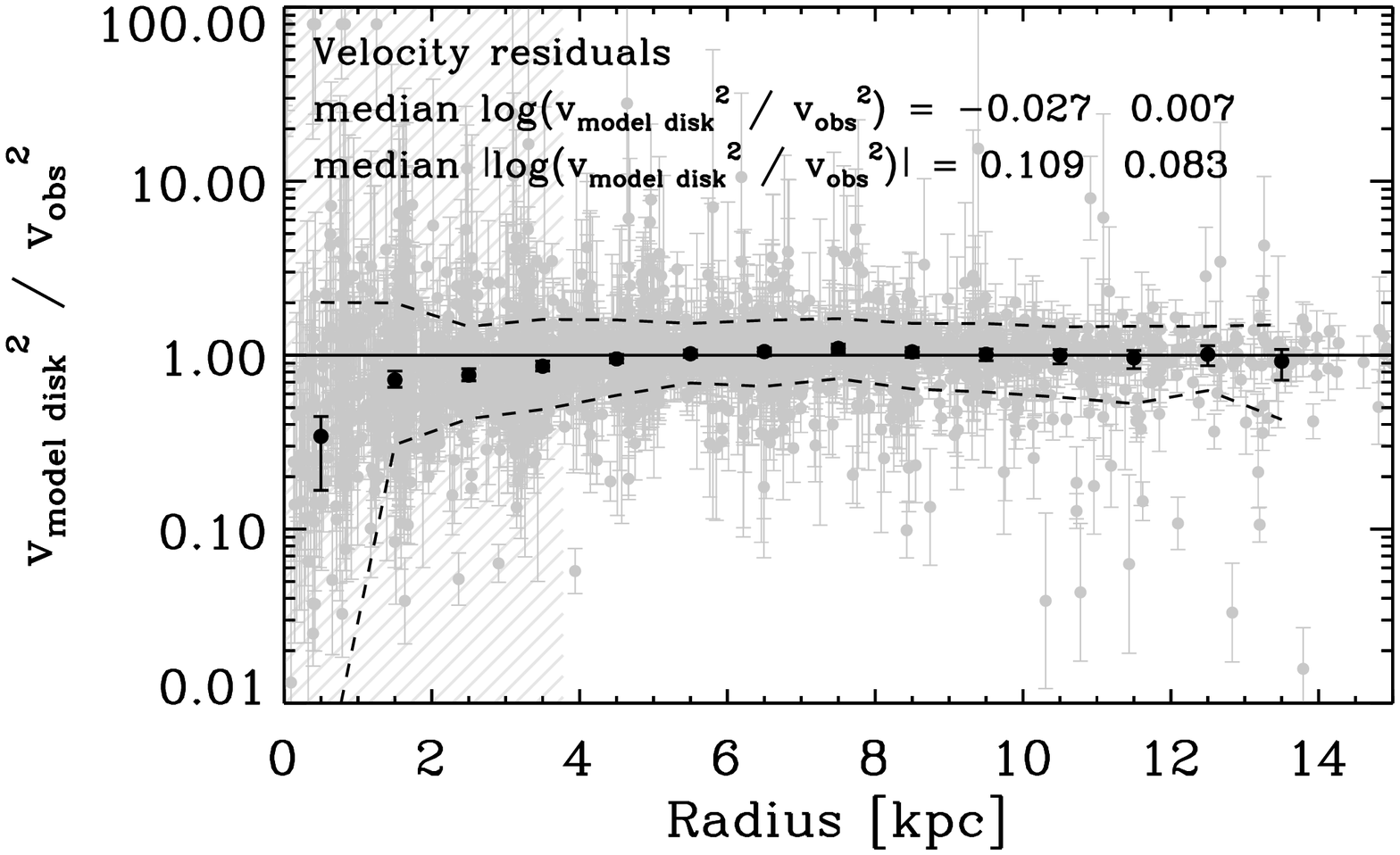}
\plotone{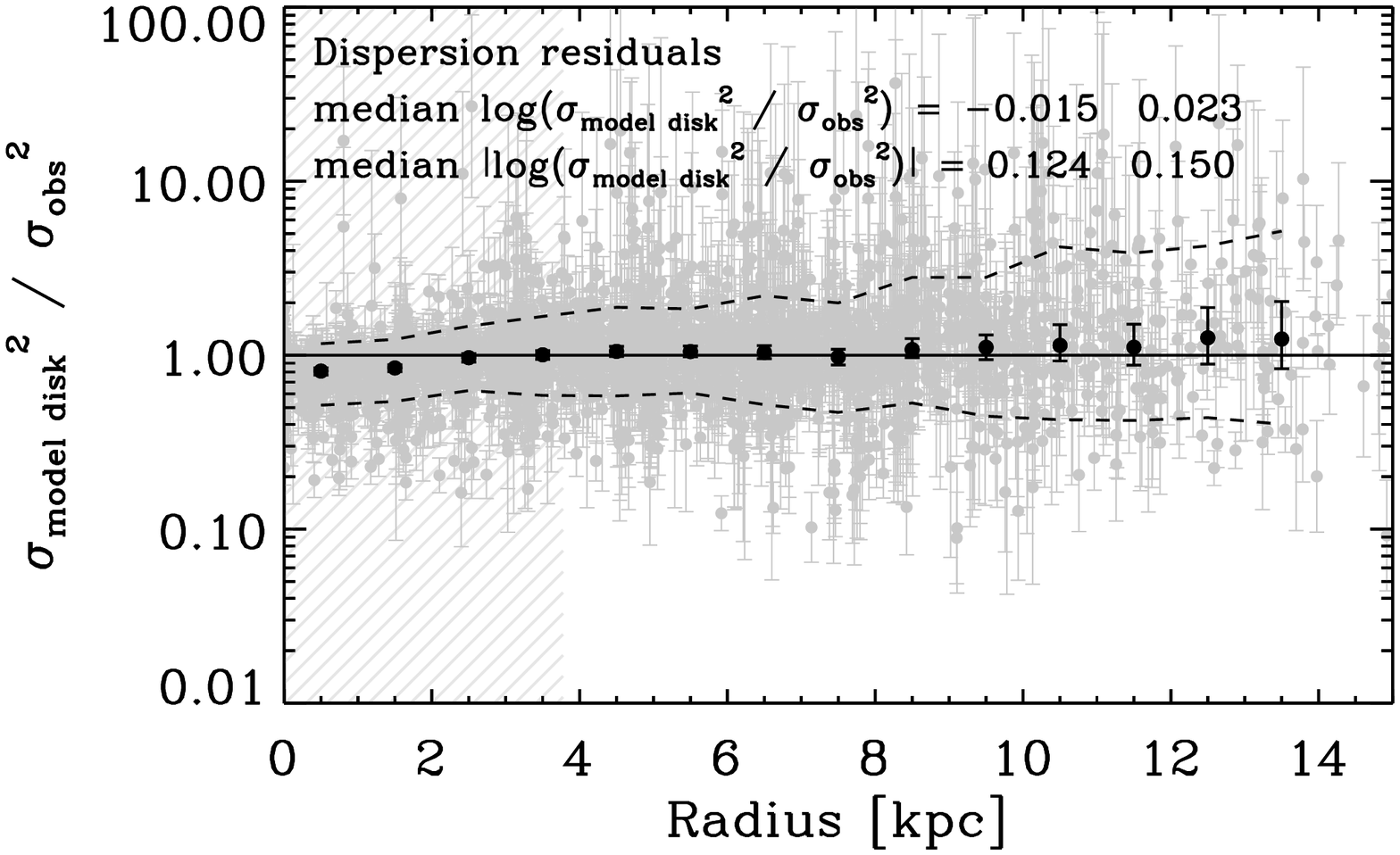}
\epsscale{1.0}
\caption{{\it Top:} Model-to-observed velocity ratio squared as a function of galactocentric radius.  The model assumes an exponential disk mass distribution.  All extracted apertures for the full sample of 240 objects are shown in gray.  Black filled circles indicate the running median, with dashed lines marking the central 68th percentile of the distribution.  The hashed region marks radii that for a typical galaxy lie less than 1 FWHM away from the center.  Median and median absolute statistics when including (left number) or excluding (right number) those apertures within $\pm 1$ FWHM from the center are listed.  {\it Bottom:} Idem, for the squared ratio of model to observed velocity dispersions as a function of radius.  No strong radial dependence of the residuals is observed.
\label{residual.fig}}
\end{figure}

\subsubsection{Fits and Residuals}
A gallery with WFC3 $H$-band and ACS $I$-band postage stamps of a subset of the galaxies in our sample are presented in Figure\ \ref{gallery.fig}, alongside their two-dimensional velocity fields, and one-dimensional velocity and velocity dispersion profiles.  Circular apertures overplotted on the velocity fields mark the extraction regions for the 1D profiles.  They often extend beyond the region for which pixel-by-pixel velocities are plotted.  This simply reflects the fact that by binning the spectra of neighboring pixels, the line emission can be traced reliably out to larger distances from the galaxy center.  In the 1D profile diagrams (two right-most panels), we also show the best-fit dynamical model.

Figure\ \ref{residual.fig} compiles the residuals between observed and modeled velocities, and between observed and modeled velocity dispersions for all 240 galaxies.  Each gray point corresponds to a radial aperture for one of the galaxies, placed at its respective galactocentric radius.  Black circles mark the running median, with error bars accounting for the oversampling that is also illustrated by the circular apertures in the middle panels of Figure\ \ref{gallery.fig}.  Dashed black lines mark the central 68th percentile.  The larger velocity residuals for the most central apertures naturally stem from the ratio of two near-zero numbers, and is restricted to the regime less than a half-light radius of the PSF away from the galaxy center.  Figure\ \ref{residual.fig} clearly demonstrates that, while we did not leave freedom in the spatial extent, inclination, or profile shape of the mass or light distribution in constructing and fitting the disk models, the resulting residuals reassuringly do not show strong trends with radial distance.

\subsection{Stellar Mass}
\label{Mstar.sec}

Stellar masses were computed by fitting stellar population synthesis (SPS) models to the spectral energy distributions of the galaxies, sampling observed-frame $U$ to 8 $\mu$m wavelengths with 16 to 43 broad and medium bands.   We used the SPS models by Bruzual \& Charlot (2003; BC03) assuming a Chabrier (2003) IMF, and followed standard procedures in the field.  Specifically, we adopted identical assumptions regarding extinction, star formation history, and metallicity as described by Wuyts et al. (2011b).  We furthermore note that, as in all our previous work, our definition of stellar mass refers to the mass in stars present at the epoch of observation (i.e., including stellar remnants, but excluding mass returned to the ISM due to stellar mass loss).

\subsection{Gas Mass}
\label{Mgas.sec}

State-of-the-art studies of the global as well as resolved Kennicutt-Schmidt relation (Kennicutt 1998) for normal main sequence SFGs at high redshift are consistent with a linear slope.  This implies that at any given time the cold gas mass and SFR are simply linked by a constant: the depletion time $t_{dep}$.  Using a combination of CO line and far-infrared continuum observations of normal SFGs over a wide redshift range ($0 < z < 3$; i.e., including the epoch of interest in this paper), Genzel et al. (2015) derived a scaling relation that describes how $t_{dep}$ depends on redshift, main sequence offset, and (to a negligible degree) stellar mass, or effectively a functional form $t_{dep}(z, SFR, M_{star})$.  By lack of individual CO measurements or fully sampled far-infrared SEDs for every galaxy in our sample, it is such a scaling\footnote{We here make use of an updated version of the Genzel et al. (2015) scaling relation, which incorporates additional data from the PHIBSS2 survey, as well as dust data by Santini et al. (2014) and Bethermin et al. (2015), and 850$\mu$m/1.2mm single band measurements analyzed with the Scoville et al. (2014) methodology.  $\log t_{dep} = a + b \log(1+z) + c \log(sSFR/sSFR_{ms,z}) + d (\log(M_{star}) - 10.5))$, where the specific SFR of the main sequence at a given redshift $sSFR_{ms,z}$ is taken from Whitaker et al. (2014), and the coefficients are $a = 0.15 \pm 0.01$, $b = -0.79 \pm 0.11$, $c = -0.43 \pm 0.01$, and $d = 0.06 \pm 0.02$ (Tacconi et al. in prep).} that we adopt to compute

\begin{equation}
M_{gas} = t_{dep}(z, SFR, M_{star}) \times SFR.
\end{equation}

Here, the SFR itself is derived from the ladder of SFR indicators introduced by Wuyts et al. (2011a).  For our specific sample, 98 galaxies have their SFR derived from the combination of rest-UV and {\it Herschel}/PACS photometry, 104 from rest-UV and {\it Spitzer}/MIPS photometry, and the remaining 38 galaxies from stellar population synthesis modeling.  The inferred molecular gas mass fractions, computed relative to the total amount of mass in baryons as $f_{gas,b} = \frac{M_{gas}}{M_{star} + M_{gas}}$, amount to $\sim36\%$ for the $z \sim 0.9$ sample, $\sim41\%$ for the $z \sim 1.5$ sample, and $\sim54\%$ for the $z \sim 2.3$ sample.

\subsection{Uncertainties}
\label{MCuncertainties.sec}

Uncertainties on the derived stellar and baryonic mass fractions were computed using 100 Monte Carlo realizations.   To this end, we perturbed the input observables (multi-wavelength photometry, size, axial ratio, velocity and dispersion measurements) within their respective formal error function for each object (see, e.g., Skelton et al. 2014; van der Wel et al. 2012), and repeated the subsequent steps in our analysis for each Monte Carlo realization: deriving the stellar mass and SFR, from it the gas mass, and fitting DYSMAL models to obtain a measure of dynamical mass.  With this approach, correlated errors are naturally taken into account.  The formal uncertainties on photometry and axial ratio of the surface brightness distribution account only for a minor contribution to the error budget.  In addition, we fold in an uncertainty associated with each of the relevant conversion steps: 0.15 dex for the SED modeling of stellar mass (characteristic for changes in adopted assumptions regarding star formation history and extinction; see, e.g., Wuyts et al. 2009), 0.25 dex for SFRs inferred from SED modeling or UV + MIPS 24$\mu$m photometry, and 0.1 dex for SFRs inferred from UV + Herschel photometry (Wuyts et al. 2011a).  On top of uncertainties in SFR and $M_{star}$, we include a 0.15 dex scatter in the $t_{dep}(z, SFR, M_{star})$ scaling (Genzel et al. 2015), and a $\pm 10^{\circ}$ error in the inclination (see, e.g., Cresci et al. 2009), even if the axial ratio of the light distribution is known to a much higher precision.  

Propagated to key quantities in our analysis, we obtain characteristic uncertainties on the dynamical mass $M_{dyn}$, the stellar mass fraction $f_{star}$ $(= M_{star}/M_{dyn})$, and the baryonic mass fraction $f_{bar}$ $(= M_{bar}/M_{dyn})$ of 0.1, 0.2 and 0.2 dex, respectively.  When calculating the median mass fraction for a set of galaxies, we determine the statistical error on the median by deriving the central 68th percentile of median values computed for each of the above Monte Carlo realizations of our sample.  The statistical errors on the median are typically small, on the order of 0.03 - 0.04 dex, meaning that systematics regarding the mass distribution (see Section\ \ref{massdistr.sec}) are not negligible in comparison.  In the remainder of the paper, when quoting or plotting errors on the median mass fraction, we simply add statistical and systematic errors in quadrature.

Systematic uncertainties in the IMF are not included in the error bars, and neither are potential contributions by other baryonic components such as atomic gas.  We discuss the latter two in Section\ \ref{discussion_fbargt1.sec}.

%%%%%
% FIG 5 [massbudget]
%%%%%
\begin {figure*}[htbp]
\epsscale{1.14}
\plotone{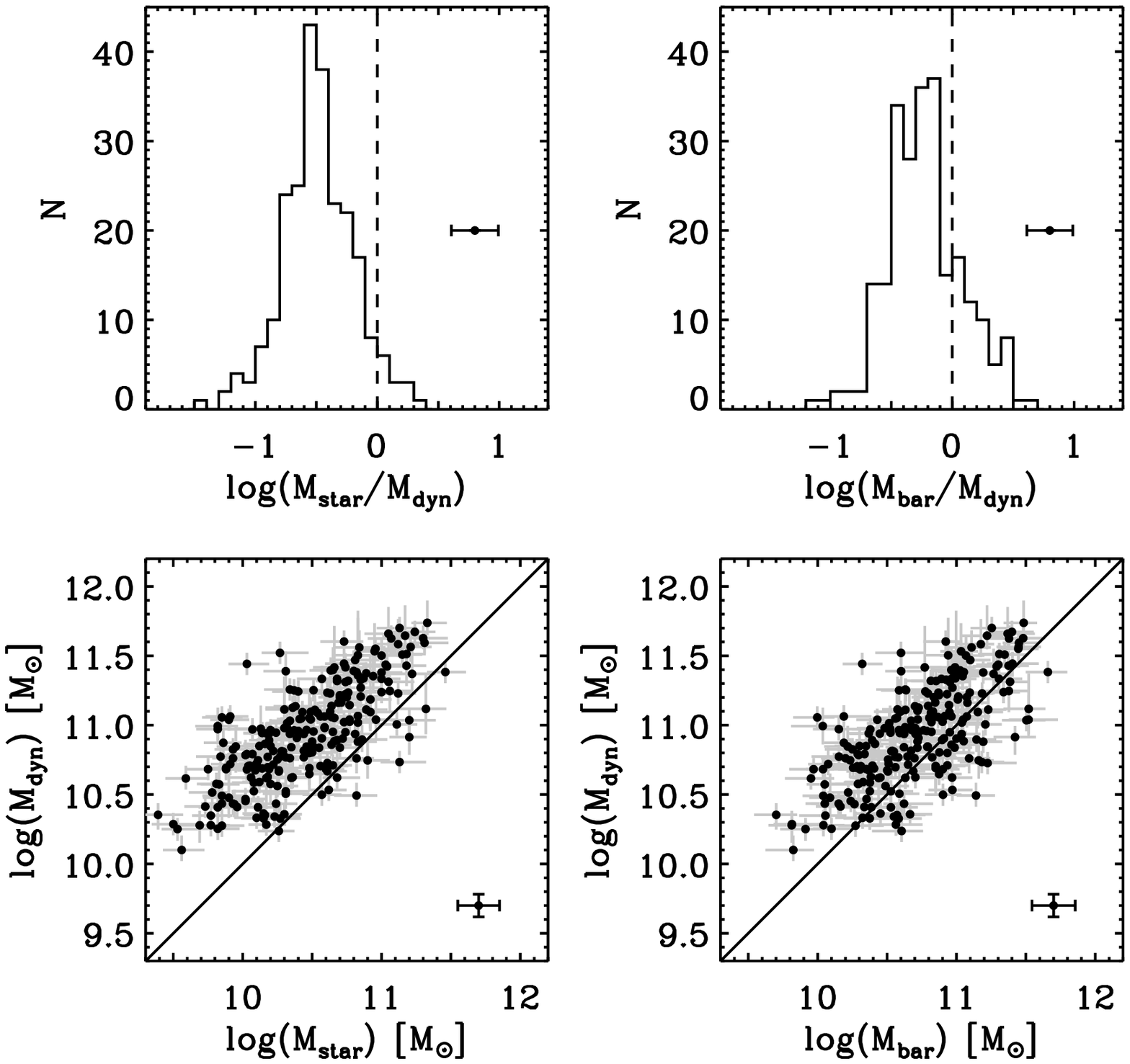}
\epsscale{1.0}
\caption{
{\it Top:} Distribution of stellar and baryonic mass fractions for the galaxies in our sample.  {\it Bottom:} Dynamical mass contrasted to the stellar mass ({\it left}) and baryonic mass ({\it right}) components.  The median error bar is indicated in the lower right corner.  The typical galaxy leaves considerable room for other mass components than stars (median $M_{star}/M_{dyn} = 0.32$).  Once accounting for the substantial gas reservoirs in high-redshift galaxies, we find the majority of disks to be baryon-dominated within their visible extent (median $M_{bar}/M_{dyn} = 0.56$).
\label{massbudget.fig}}
\vspace{-0.5cm}
\end{figure*}

\section {Results}
\label{results.sec}

\subsection{The Mass Budget in Early Star-forming Disks}
\label{massbudget.sec}

We consider the distribution of stellar and baryonic mass fractions in Figure\ \ref{massbudget.fig}.  Clearly, not all galaxies exhibit the same breakdown in their mass budget.  The central 68th percentile intervals of the respective distributions are $\log(f_{star})_{68} \sim [-0.75;\ -0.21]$ and $\log(f_{bar})_{68} \sim [-0.49;\ 0.09]$.  The total range exceeds an order of magnitude when including the extremes.  This observed range cannot be accounted for by our formal uncertainty estimates solely, and hence has to reflect variations in the intrinsic physical properties among galaxies.  While on average stars only account for about a third of the total mass (median $f_{star} = 0.32_{-0.07}^{+0.08}$), the mass budget after adding the substantial gas reservoirs implied by CO and dust scaling relations is typically baryon-dominated (median $f_{bar} = 0.56_{-0.12}^{+0.17}$, where the error reflects the total error in the median rather than the width of the overall $f_{bar}$ distribution; see Section\ \ref{MCuncertainties.sec}).

Cases with $f_{star} > 1$ are rare, amounting to 5\% of the total sample.  None of these objects have $M_{star}$ exceeding $M_{dyn}$ by more than 2$\sigma$.  The same is not true when evaluating $f_{bar}$.  Baryonic mass fractions in the unphysical regime (i.e., $f_{bar} > 1$) are found for 23\% of the galaxies in our sample, although this fraction reduces to 11/3/1\% when requiring a deviation from the physical limit of unity at the 1/2/3$\sigma$ level.

We conclude from the comparison of stellar and dynamical masses that, reassuringly, there is significant room for other mass components than stars.  Adding molecular gas reveals the baryon-dominated nature of most galaxies, although at face value indications of missing mass remain in three quarters of the sample.  This finding is consistent with and improves on recent results on high-z SFGs showing they are baryon-dominated within $R_e$ (F\"{o}rster Schreiber et al. 2009; Burkert et al. 2016; Price et al. 2016; Stott et al. 2016).

The bottom left panel of Figure\ \ref{massbudget.fig} illustrates that our independent measurements of stellar and dynamical mass show a highly significant correlation.  The same is true for the relation between dynamical and baryonic mass.  Taking out the overall trend that more massive SFGs contain more mass in all mass components, we find evidence for systematic dependencies of $f_{star}$ and $f_{bar}$ on various, often interconnected, galaxy properties, most notably surface density.  We discuss these in Section\ \ref{dependence.sec}, but first demonstrate that inclination uncertainties cannot account for the aforementioned missing mass.

\subsection {A Statistical Perspective on the Inclination Distribution of the Galaxy Ensemble}
\label{incdistr.sec}

%%%%%
% FIG 6 [inclination distribution]
%%%%%
\begin {figure}[t]
\epsscale{1.14}
\plotone{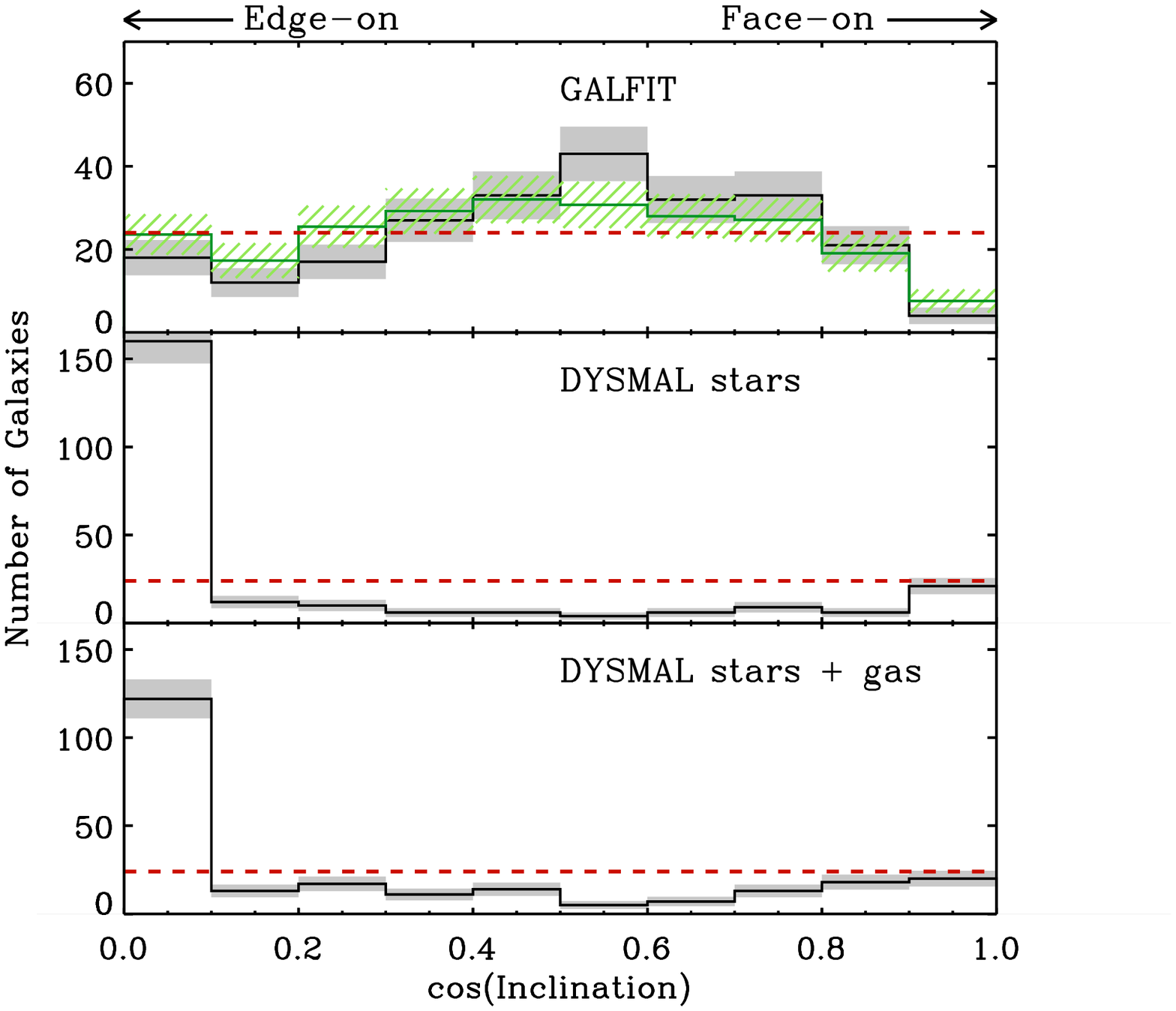}
\epsscale{1.0}
\caption{Inclination distribution of our KMOS$^{3D}$ sample (black) and the complete underlying 3D-HST population matched in mass, redshift, size and star formation activity (green), contrasted to the expectation for random viewing angles (dashed red line).  {\it Top panel:} Inclinations as inferred from the axial ratio $b/a$ measured with GALFIT (equation\ \ref{inc.eq}).  {\it Middle panel:} Distribution of inclinations that yield the best fit of the DYSMAL stellar mass models to the observed kinematics.  {\it Bottom panel:} Distribution of inclinations that yield the best fit of the DYSMAL stellar + gas mass models to the observed kinematics.  Our KMOS sample shows a similarly flat distribution of inclinations as the underlying matched population, with only a slight paucity of very edge- or face-on systems compared to the expectation for random viewing angles.  In contrast, an implausibly large number of edge-on inclinations would be required to optimally reproduce the observed kinematics when fixing the total mass to our best estimates of the stellar or baryonic mass present.
\vspace{-0.5cm}
\label{inclination.fig}}
\end{figure}

Thus far, we chose to fix the disk inclination to the value informed by HST imagery when fitting for $M_{dyn}$.  Rather than leaving both mass and inclination free in the fit, and opening ourselves to well-known degeneracies that are hard to break with KMOS data alone (for more discussion in the context of high-z galaxies, see, e.g., Cresci et al. 2009; Bouch\'{e} et al. 2015), we now take the reverse approach of fixing the mass to $M_{star}$ or $M_{bar}$ and fitting for the inclination that best describes the observed kinematics.  This allows us to address the following question: could it be that in reality stars (or baryons, as gas is undoubtedly present) make up all of the mass, but that inclination uncertainties mistakenly led us to infer the presence of 'missing mass'?

On an individual object basis, we find that in the majority of cases (but not all) an acceptable fit to the observed kinematics can be obtained when fixing $M_{dyn} \equiv M_{bar}$ and relaxing all constraints on inclination.  For the exceptions, amounting to 10\% of the sample, even a dynamical mass estimate based on the assumption of an edge-on viewing angle yields $M_{bar} < M_{dyn}$ at the 3$\sigma$ level.

However, the large size of our sample allows us to go beyond considerations at the individual object level, and carry out an investigation for the galaxy ensemble as a whole.  Namely, whether the {\it distribution} of inclinations that, given our baryonic mass models, yield the best fit to the observed rotation curves is consistent with the statistical expectation for random viewing angles.  For an ensemble of galaxies with random orientations with respect to an observer, $\cos(i)$ is uniformly distributed between 0 and 1 (Rix et al. 1997).  Before assessing whether this is the case for the inclination values that yield the best possible fits to the observed velocity and dispersion profiles, it is worth testing that the assumption of random viewing angles is at all applicable to our sample.  After all, disk galaxies contain dust, and their emission may be attenuated by a thicker column of obscuring material when seen edge-on.  This could reduce the H$\alpha$ signal-to-noise ratio, and potentially cause them to drop out of our sample more easily.  On the other hand, as reported in Section\ \ref{sample.sec}, some galaxies, while H$\alpha$ detected, did not pass our sample selection because their face-on view prevented meaningful constraints on the enclosed dynamical mass.  The top panel of Figure\ \ref{inclination.fig} illustrates that nevertheless the inclination distribution of our sample as inferred from axial ratio measurements on the HST imaging is relatively flat, mimicking that of the underlying matched 3D-HST population (normalized distribution shown in green).  Modulo a minor dearth of very edge-on and face-on systems the distributions are in line with the expectation for random viewing angles.

The middle and bottom panel of Figure\ \ref{inclination.fig} convincingly demonstrate that this is no longer the case when considering the best-fit inclinations from kinematic models where we fixed the mass to the stellar or baryonic mass, respectively.  Leaving no freedom to $M_{dyn}$, very extreme inclinations frequently need to be invoked to yield the best possible description of the observed kinematics.  The distributions particularly show a strong peak in the highest inclination (near edge-on) bin.  Clearly, these orientations are overrepresented when attempting to reproduce the amplitude of velocity gradients without additional mass components.

%%%%%%%%
% TABLE 1
%%%%%%%%
\begin {deluxetable*}{lrrr}
\tablecolumns{4}
\tablewidth{\linewidth}
\tablecaption{Stellar and baryonic mass fractions of KMOS$^{\rm 3D}$ star-forming disk galaxies, and estimates for a mass-complete sample of SFGs above a fixed or evolving mass limit (see text). \label{massfrac.tab}}
\tablehead{
\colhead{Redshift} & \colhead{$\log(M_{star}/M_{dyn})$\tablenotemark{a}} & \colhead{$\log(M_{bar}/M_{dyn})$\tablenotemark{a}} & \colhead{$\log(M_{bar, K98}/M_{dyn})$\tablenotemark{a,b}}
}
\startdata
\multicolumn{4}{c}{KMOS$^{\rm 3D}$ sample} \\
\cline{1-4}
\multicolumn{4}{c}{} \\
$0.6 < z < 1.1$ & $-0.55 \pm 0.12$ [-0.80; -0.34] & $-0.35 \pm 0.09$ [-0.60; -0.16] & $-0.32 \pm 0.09$ [-0.50; -0.09] \\
$1.3 < z < 1.7$ & $-0.51 \pm 0.11$ [-0.75; -0.26] & $-0.27 \pm 0.11$ [-0.53; -0.03] & $-0.21 \pm 0.11$ [-0.43; -0.03] \\
$2.0 < z < 2.6$ & $-0.38 \pm 0.11$ [-0.69; -0.10] & $-0.05 \pm 0.14$ [-0.34; 0.24] & $-0.04 \pm 0.14$ [-0.26; 0.26] \\
$1.3 < z < 2.6$ & $-0.43 \pm 0.11$ [-0.71; -0.14] & $-0.14 \pm 0.13$ [-0.40; 0.21] & $-0.11 \pm 0.13$ [-0.32; 0.20] \\
All             & $-0.50 \pm 0.11$ [-0.75; -0.21] & $-0.25 \pm 0.11$ [-0.49; 0.09] & $-0.19 \pm 0.11$ [-0.43; 0.12] \\
\multicolumn{4}{c}{} \\
\multicolumn{4}{c}{Mass-complete star-forming population with $\log(M_{star}) > 9.8$} \\
\cline{1-4}
\multicolumn{4}{c}{} \\
$0.6 < z < 1.1$ & $-0.58 \pm 0.12$ [-0.81; -0.38] & $-0.39 \pm 0.10$ [-0.64; -0.18] & $-0.36 \pm 0.10$ [-0.55; -0.13] \\
$1.3 < z < 1.7$ & $-0.50 \pm 0.12$ [-0.75; -0.18] & $-0.28 \pm 0.12$ [-0.52; 0.07] & $-0.19 \pm 0.11$ [-0.45; 0.09] \\
$2.0 < z < 2.6$ & $-0.49 \pm 0.11$ [-0.75; -0.17] & $-0.14 \pm 0.15$ [-0.40; 0.22] & $-0.11 \pm 0.14$ [-0.33; 0.17] \\
\multicolumn{4}{c}{} \\
\multicolumn{4}{c}{Progenitors of $\log(M_{star,\ z \sim 0}) > 10.7$} \\
\cline{1-4}
\multicolumn{4}{c}{} \\
$0.6 < z < 1.1$ & $-0.45 \pm 0.13$ [-0.60; -0.24] & $-0.32 \pm 0.11$ [-0.44; -0.13] & $-0.29 \pm 0.10$ [-0.40; -0.13] \\
$1.3 < z < 1.7$ & $-0.42 \pm 0.12$ [-0.63; -0.25] & $-0.20 \pm 0.11$ [-0.37; -0.05] & $-0.23 \pm 0.11$ [-0.33; -0.06] \\
$2.0 < z < 2.6$ & $-0.37 \pm 0.11$ [-0.68; -0.08] & $-0.06 \pm 0.14$ [-0.35; 0.25] & $-0.06 \pm 0.14$ [-0.25; 0.24] \\
%\vspace{-2mm}
\enddata

\tablenotetext{a}{\footnotesize Median, error on the median, and between brackets the associated central 68th percentile range.}
\tablenotetext{b}{\footnotesize Baryonic mass fraction based on gas masses computed following the inverse Kennicutt (1998) relation.}
\end{deluxetable*}
%XXX

We conclude that, while inclination uncertainties undoubtedly affect the assessment of the mass budget breakdown in individual galaxies, they cannot account for the observed missing mass in the ensemble of galaxies (i.e., the fact that on average $M_{dyn} > M_{bar}$ and $M_{dyn} \gg M_{star}$).

\vspace{3mm}

\subsection{Trends with Other Galaxy Properties}
\label{dependence.sec}

In Section\ \ref{massbudget.sec}, we found that distant galaxies feature a broad range of stellar-to-dynamical and baryonic-to-dynamical mass fractions.  Here, we investigate whether the observed variations in mass fraction correlate with other galaxy properties.

\subsubsection{Redshift Dependence}
\label{redshift.sec}

\begin {figure}[t]
\plotone{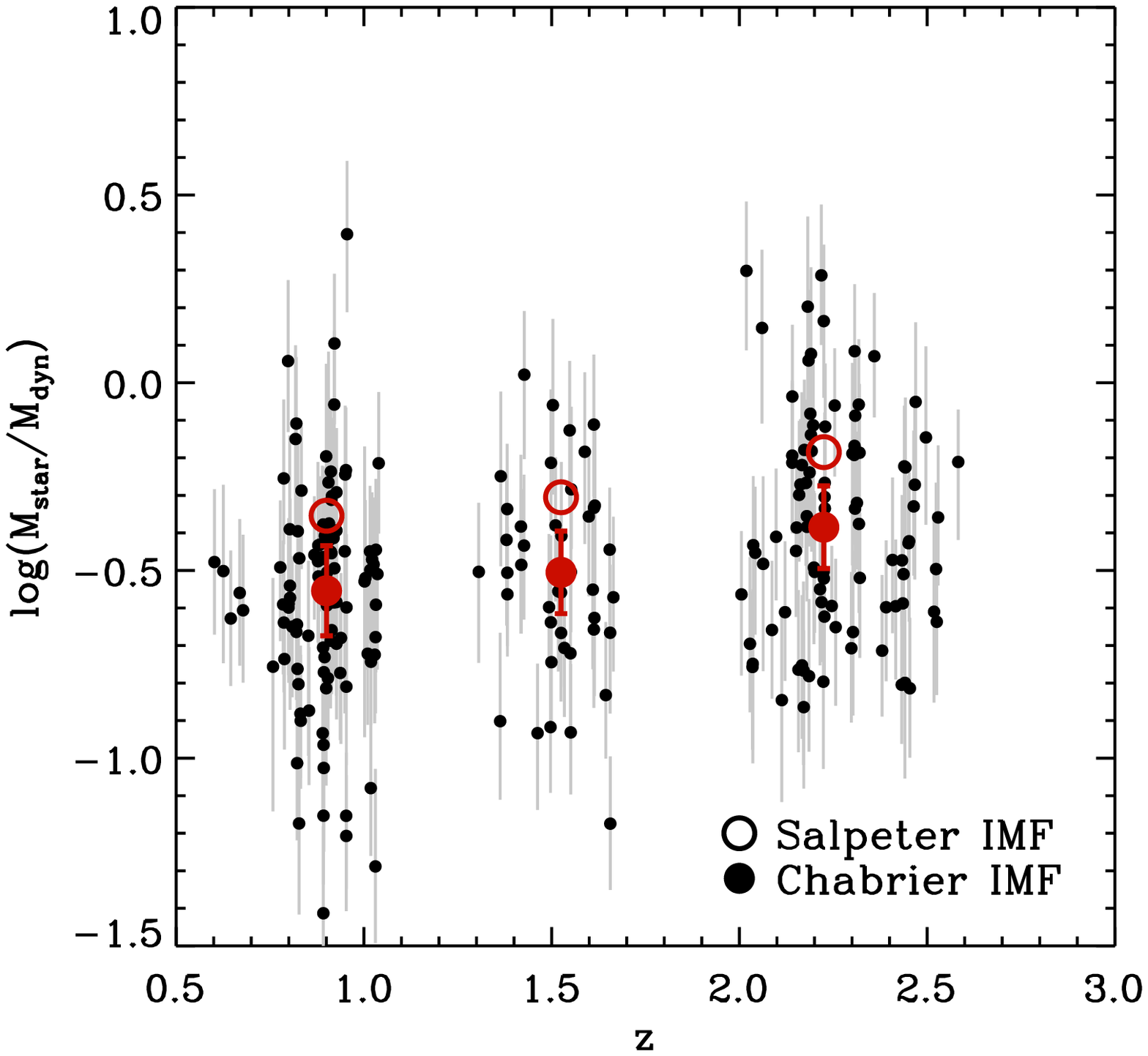}
\plotone{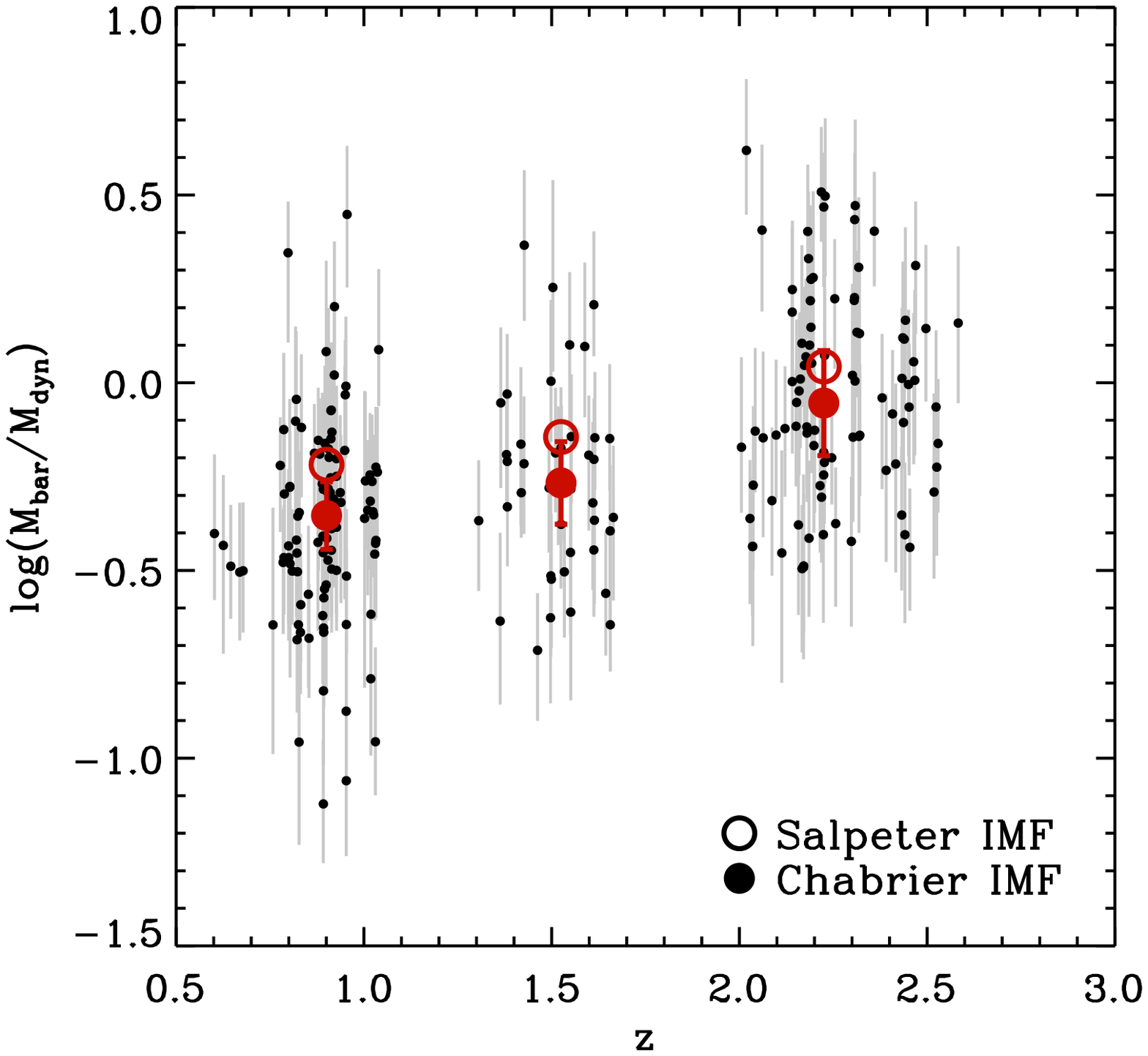}
\caption{Stellar mass fraction ({\it top row}) and baryonic mass fraction ({\it bottom row}) as a function of redshift.  Large circles mark the median for each redshift bin.  Filled circles represent results assuming the default Chabrier (2003) IMF, while large empty circles indicate the shift if a Salpeter (1955) IMF were adopted instead.  Galaxies in our highest redshift bin ($z \sim 2.3$) are entirely baryon dominated within $R_e$.
\label{zevol.fig}}
\end{figure}

Combining $YJ$, $H$ and $K_s$ observations, our sample spans a wide dynamic range in redshift ($0.6 < z < 2.6$), sampling as much as 40\% of the history of the universe.  It is thus natural to consider whether the breakdown of the mass budget evolves over the different epochs probed.  We investigate this in Figure\ \ref{zevol.fig}, with Table\ \ref{massfrac.tab} summarising the median mass fractions and scatter in different redshift intervals.  A modest increase in the median stellar mass fraction by a factor 1.5 is noted between the lowest ($z \sim 0.9$) and highest ($z \sim 2.3$) redshift bins, an amount that is smaller than the typical 0.25 - 0.3 dex scatter observed within each bin.  At face value, the sign of this offset ($f_{star}$ increasing with redshift) is somewhat counterintuitive, given studies of molecular gas reservoirs and their evolution over cosmic time (e.g., Genzel et al. 2015 and references therein).  In the absence of other mass components, or if enclosed dark matter fractions remain constant over time, declining baryonic gas fractions would lead to rising $f_{star}$ as time proceeds.  Clearly, this is not observed.  Within the $z \sim 0.9$ bin, more gas-rich galaxies on average feature lower $f_{star}$, but no such trend is significantly present at higher redshifts, or when considering the full sample.  If galaxies in our low-redshift bin on average feature higher dark matter fractions within $R_e$ than those in the high-redshift bin, this could explain the observed trend.  We return to this interpretation in Section\ \ref{surfdens.sec}, and within the context of cosmological simulations in Section\ \ref{discussion_illustris.sec}.

The empirical gas scaling relations we adopt suggest a decline in gas fractions for the galaxies in our sample from $\sim 54\%$ for the highest redshift bin to $\sim 36\%$ for the lowest redshift bin.  Consequently, the modest trend of increasing mass fraction with redshift is enhanced when adding the gas reservoirs and considering the full contribution by baryons to the mass budget (Figure\ \ref{zevol.fig}, lower panel).  It is evident that star-forming disk galaxies above $z \sim 2$ with sizes $R_e \gtrsim 2$ kpc are heavily baryon dominated.  This strenghtens the earlier assessment of F\"{o}rster Schreiber et al. (2009), is in agreement with Burkert et al. (2016) and slit-based spectroscopic measurements by Price et al. (2016), and unlike local disk galaxies which feature dark matter fractions of around $\sim 50\%$ within 2.2 disk scale lengths (Courteau \& Dutton 2015).  Adopting a Salpeter (1955) rather than Chabrier (2003) IMF enhances the tension with dynamical constraints slightly (see also Price et al. 2016), although it should be noted that the impact of this assumption on the inferred baryonic mass fraction is reduced for galaxies where molecular gas contributes most of the baryons.

Analyzing KMOS observations of $z=0.8 -1$ galaxies spanning a similar radial range as available for our sample (typically out to 9-10 kpc), Stott et al. (2016; KROSS) recently reported stellar mass fractions of $\log(M_{star}/M_{dyn}) \approx -0.66$ and baryonic mass fractions of $\log(M_{bar}/M_{dyn}) \approx -0.40$.  At higher redshifts ($1.4 \leq z \leq 2.6$), Price et al. (2016; MOSDEF) modeled slit-based observations with MOSFIRE, finding the mass budget within $R_e$ to break down as $\log(M_{star}/M_{dyn}) \approx -0.36$ and $\log(M_{bar}/M_{dyn}) \approx -0.04$.  These numbers agree within 1 sigma with the values tabulated in Table\ \ref{massfrac.tab}.  We note that the precise numerical comparison between these survey results should not be overinterpreted, for three reasons.  First, both the KROSS and MOSDEF samples extend down to lower masses than considered here, with a median stellar mass of $\langle \log(M_{star}) \rangle=10$ compared to $\langle \log(M_{star}) \rangle=10.5$ for our sample.  Second, methodologies differ.  Stott et al. (2016), for example, assume a spheroidal mass distribution and do not account for a degree of pressure support in deriving $M_{dyn}$.  While, judging from their analysis, imposing a higher stellar mass cut would increase the characteristic $f_{star}$ of the sample, including a contribution of pressure support in the dynamical modeling would have the compensating effect of increasing $M_{dyn}$ and hence yielding a lower $f_{star}$.  Price et al. (2016) on the other hand rely for 80\% of their sample on virial estimates using unresolved galaxy-integrated velocity dispersions.  Finally, both studies employ the (inverse) Kennicutt (1998) relation to translate the observed surface density of star formation to a gas surface density and subsequently integrated $M_{gas}$.  To illustrate the impact of the latter assumption relative to the Genzel et al. (2015) gas scaling relations adopted thus far, we include for reference in Table\ \ref{massfrac.tab} baryonic mass fractions computed under the assumption of the Kennicutt (1998) relation.

%%%%
% FIG 7 [redshift]
%%%%%

%%%%%
% FIG 8 [redshift underlying pop]
%%%%%
\begin {figure}[t]
\epsscale{1.14}
\plotone{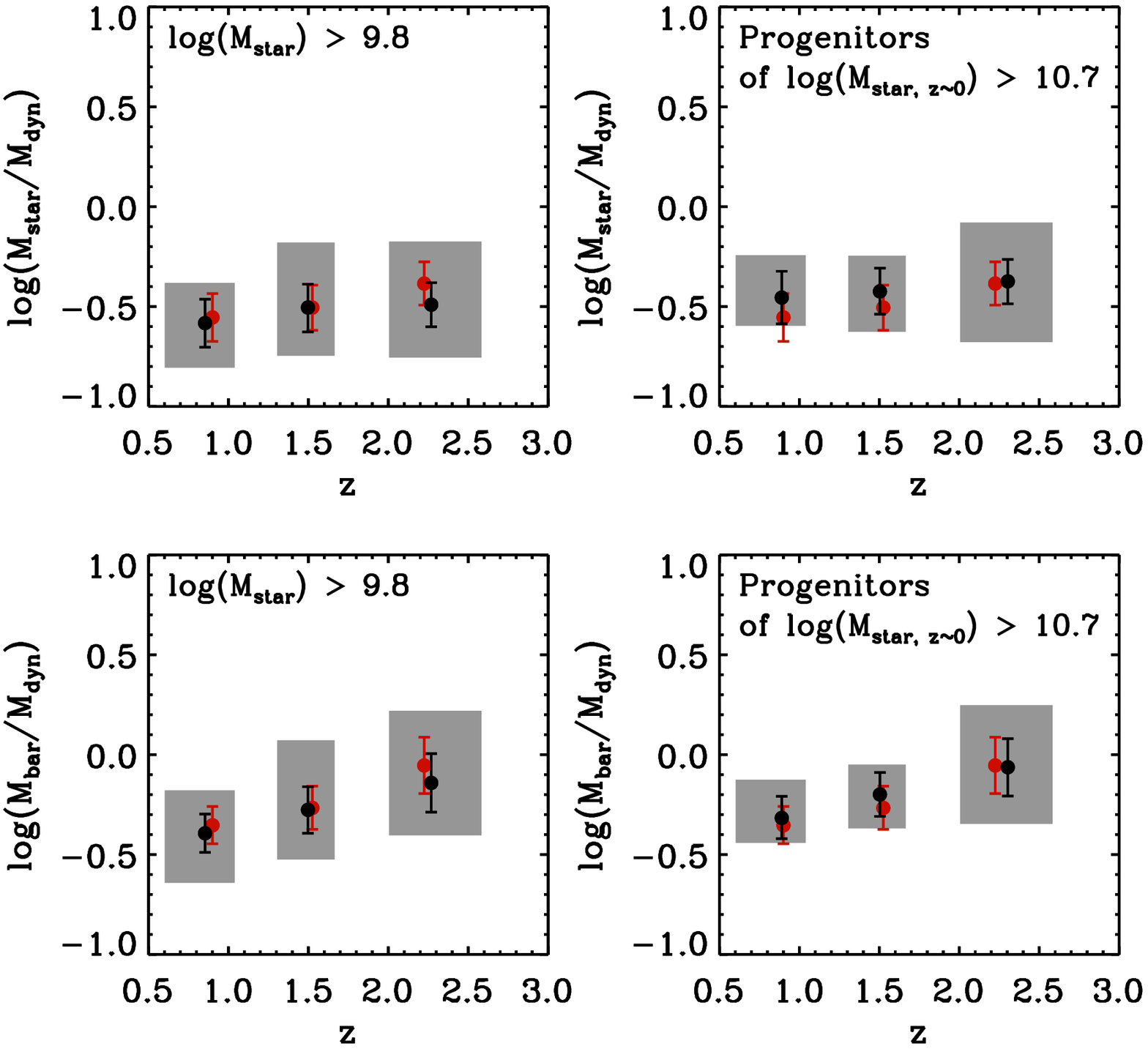}
\epsscale{1}
\caption{Redshift evolution of the stellar and baryonic mass fraction with weights applied to the galaxies in our kinematic sample to represent as closely as possible the galaxy population above a fixed mass limit ({\it left panels}), or a mass limit that is shifting with redshift to trace at each epoch the progenitors of $\log(M_{star}) > 10.7$ galaxies in the present-day universe ({\it right panels}).  Central 68th percentiles of the distribution in each redshift bin are marked with grey rectangles.  Median mass fractions of the actual (unweighted) KMOS$^{\rm 3D}$ sample are shown for reference in red, taken from Figure\ \ref{zevol.fig}.
\label{zevol_parent.fig}}
\end{figure}

As detailed in Section\ \ref{sample.sec}, our kinematic sample shares many similarities with the underlying population of SFGs, but is statistically inconsistent with being drawn randomly from its mass distribution.  In principle, if $f_{star}$ or $f_{bar}$ were strongly mass dependent quantities, the flatter mass distribution of the kinematic sample could imply that the redshift evolution for a sample that is complete down to $10^{9.8}\ M_{\sun}$ may look different.  We investigate this in Figure\ \ref{zevol_parent.fig} using the following crude approach.  To each galaxy from the mass-complete 3D-HST sample described in Section\ \ref{sample.sec} we assign a value of $f_{star}$ and $f_{bar}$ based on the KMOS$^{\rm 3D}$ galaxy that resembles it most closely in intrinsic properties (position in SFR - mass space and surface density).  Or in other words, we effectively assign weights to each KMOS$^{\rm 3D}$ galaxy in order for the sample to better mimic the underlying population.  Figure\ \ref{zevol_parent.fig} exhibits trends that are familiar from our investigation of the kinematic sample itself, with no significant redshift evolution in stellar mass fractions (formally a factor of 1.2), but rising $f_{bar}$ with redshift, reaching the heavily baryon-dominated regime at $z > 2$.

%%%%%
% FIG 9 [surfdens]
%%%%%
\begin {figure*}[htbp]
\plotone{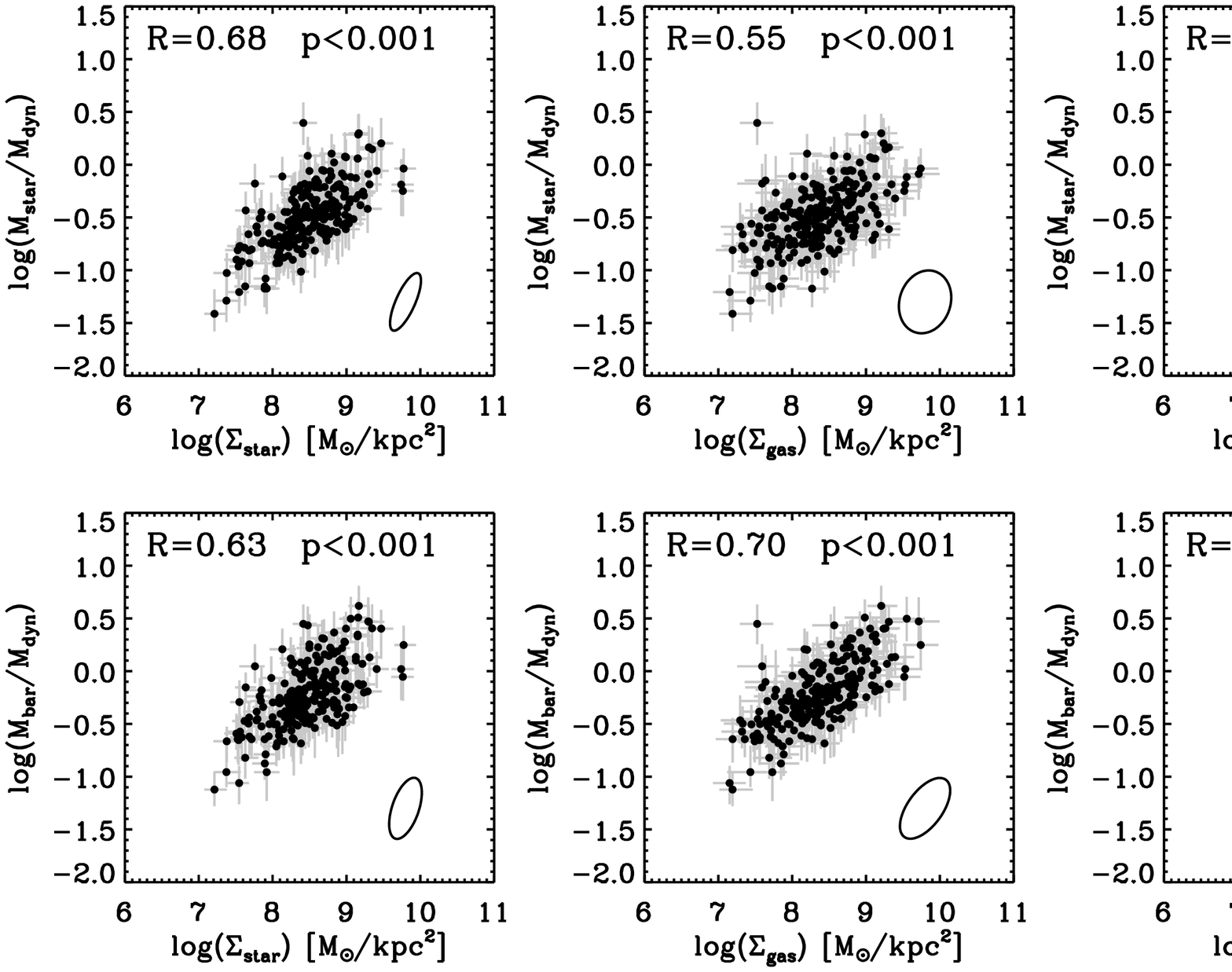}
\caption{Dependence of stellar mass fraction ({\it top row}) and baryonic mass fraction ({\it bottom row}) on surface density of stars ({\it left panels}), gas ({\it middle panels}), and the sum of both baryonic components ({\it right panels}).  Characteristic error ellipses, correlation coefficients and p values are marked in each panel.  The derived mass fractions correlate significantly with surface density, increasing from the most extended to the more compact systems.
\label{corr_surfdens.fig}}
\end{figure*}

It is worth noting that by selecting galaxies above a fixed mass limit, one does not trace progenitor - descendant populations across cosmic time.  Galaxies grow in mass, and new galaxies will hence cross the mass limit and enter the sample as time proceeds.  Selecting galaxies above a mass limit that evolves with redshift, corresponding to a fixed cumulative comoving number density is a commonly adopted alternative (e.g., van Dokkum et al. 2010, 2013, 2015; Papovich et al. 2011; Patel et al. 2013).  Although not without shortcomings (see, e.g., Torrey et al. 2015), this approach comes closer to tracing how individual galaxies evolve through cosmic time.  The right-hand panels of Figure\ \ref{zevol_parent.fig} explore the evolution in mass fractions inferred this way for progenitors of galaxies with $\log(M_{star,\ z \sim 0}) > 10.7$ today, where the evolving mass limit was derived from the galaxy stellar mass functions by Tomczak et al. (2014) and is illustrated in the middle panel of Figure\ \ref{sample.fig}.  Again, very similar trends emerge.  We infer stellar mass fractions to be relatively constant over the time interval studied, leaving substantial room for other mass components, and baryonic mass fractions that were higher at earlier epochs.  The statistics on inferred mass fractions above a fixed or evolving mass limit are also listed in Table\ \ref{massfrac.tab}.

Aside from redshift, we explored correlations of the stellar and baryonic mass fractions with various other parameters, including different mass components, gas fractions, specific star formation rates, and galaxy sizes.  By far the strongest correlations are found with measures of surface density.  Given the size evolution of galaxies with cosmic time (e.g., van der Wel et al. 2014), it is entirely plausible that the redshift trends described in this Section arise at least in part as an indirect imprint of such an underlying relation with surface density.

\subsubsection{Surface Density Dependence}
\label{surfdens.sec}

Figure\ \ref{corr_surfdens.fig} contrasts the stellar and baryonic mass fractions to the surface density of different mass components.  The strongest correlations ($R > 0.68$) are observed between $f_{star}$ and $\Sigma_{star}$ on the one hand, and $f_{bar}$ and $\Sigma_{gas}$ (or $\Sigma_{bar}$) on the other hand:

\vspace{-2mm}
\begin{eqnarray}
\label{massfrac_surfdens.eq}
%\begin{tabular}{llrl}
%$\log(f_{star})$ & = & $(-4.29 \pm 0.09) + (0.44 \pm 0.03) \log(\Sigma_{star})$ & $ \ $  \vspace{1.5mm} \\
%$\log(f_{bar})$ & = & $(-4.04 \pm 0.08) + (0.45 \pm 0.03) \log(\Sigma_{gas})$ & $ \ $ \vspace{1.5mm} \\ 
%$\log(f_{bar})$ & = & $(-4.49 \pm 0.09) + (0.49 \pm 0.03) \log(\Sigma_{bar})$ & $ \ $ \vspace{1.5mm}
%\end{tabular}
\noindent \log(f_{star}) & = & -0.47 + 0.49 [\log(\Sigma_{star}) - 8.5]\ \ \ \ \label{eqn_1} \\
\noindent \log(f_{bar}) & = & -0.14 + 0.52 [\log(\Sigma_{gas}) - 8.5] \label{eqn_2} \\
\noindent \log(f_{bar}) & = & -0.34 + 0.51 [\log(\Sigma_{bar}) - 8.5] \label{eqn_3}
\end{eqnarray}

with uncertainties on the intercept and slope of 0.01 and 0.03, respectively.  As the error ellipses illustrate, uncertainties along both axes are not independent given shared information, but the dynamic range of the observed correlations significantly exceeds what is expected from correlated uncertainties, boosting confidence in the physical reality of the observed trends.  It is noteworthy that the correlations with surface density are stronger than if one considers the total mass of the respective components.

At stellar surface densities of around $\Sigma_{star} \sim 10^9\ M_{\sun}/kpc^2$, most of the dynamical mass can be accounted for by stars.  This reduces to of order 10\% when considering the subset of galaxies with the most diffuse stellar distributions ($\Sigma_{star} < 10^8\ M_{\sun}/kpc^2$).  The observed trend echos findings by Barro et al. (2014), who reported higher stellar mass fractions for their sample of 13 compact SFGs than present in a non-compact SFG reference sample.  Burkert et al. (2016) studied the angular momentum parameter distribution in high-redshift galaxies, and also found the most significant (negative) correlation to be with stellar surface density (computed within the half-light radius, as we do here).

Similar to the $f_{star} - \Sigma_{star}$ relation, our analysis implies that baryons contribute a progressively larger fraction of the total mass budget as gas and/or baryonic surface densities increase.  If we were to adopt the inverted Kennicutt (1998) relation to derive gas masses, a significant positive correlation between $f_{bar}$ and $\Sigma_{gas/bar}$ remains, albeit with a slightly reduced Pearson correlation coefficient (R = 0.63).  The $f_{star} - \Sigma_{star}$ relation obviously remains unaffected.

A plausible interpretation of the observed relations is that for high surface density systems, the visible extent of the galaxy traced by the H$\alpha$ and $H$-band emission probes mostly the inner, baryon-dominated part of the halo.  More diffuse, lower surface density systems will probe further into the halo, where dark matter makes up a larger fraction of the mass budget.

\section{Discussion}
\label{discussion.sec}

\subsection{Objects with $f_{bar} > 1$}
\label{discussion_fbargt1.sec}

In the previous Section, we described a picture of baryonic distributions of varying size embedded in larger scale dark matter halos.  While this can at least in qualitative terms explain the presence of a relation between the surface density and the baryonic mass fraction within the visible extent of our galaxies, it can by itself not be responsible for the $f_{bar} > 1$ values seen at high inferred gas surface densities.  Those are by definition unphysical.  At $\Sigma_{gas} \gtrsim 10^9\ M_{\sun}/kpc^2$, such cases make up 70\% of the galaxies.  Here, we briefly discuss which factors, other than random uncertainties scattering baryon-dominated sources above the physical limit, may contribute to their presence.

\subsubsection{Stellar masses}

Stellar masses computed through SED modeling are subject to several potential systematic uncertainties (see, e.g., Wuyts et al. 2009).  However, most of these have the tendency to lead to underestimates rather than overestimates of the true mass present.  E.g., if the true IMF follows a Salpeter (1955) rather than the adopted Chabrier (2003) IMF, if the true star formation history features bursts on top of an underlying older stellar population, or if the dust distribution is such that the total amount of attenuation could not be recovered.  In these scenarios, the tension with dynamical constraints would be amplified instead of remedied.  Some IMF studies of nearby early-type galaxies with $\sigma > 200\ km/s$ favor an even more bottom-heavy IMF than Salpeter, with mass-to-light ratios being larger by up to a factor of 2 (Spiniello et al. 2012; Conroy \& van Dokkum 2012; Ferreras et al. 2013).  If the stars in those galaxies were formed at the redshifts and in the type of galaxies contained within our KMOS$^{\rm 3D}$ sample, this would naturally enhance tensions with dynamical constraints further.  SPS models by Maraston (2005) instead of BC03 would reduce the stellar masses of the galaxies in our sample by a factor 1.4.  We note, however, that recent spectroscopic (Zibetti et al. 2013) and spectro-photometric (Kriek et al. 2010) studies of post-starburst galaxies seem to favor BC03 models, in that they exhibit relatively low rest-frame near-infrared luminosities and lack the prominent CO bandheads characteristic for the M05 models in that wavelength regime (although see also Capozzi et al. 2016 for an opposing view).

\subsubsection{Gas masses}

Turning to the gaseous mass component, we remind the reader that we only accounted for gas in the molecular phase.  Atomic (HI) gas columns in nearby galaxies have been shown to saturate at around $\sim 10\ M_{\sun}\ pc^{-2}$ (Bigiel \& Blitz 2012), corresponding to the threshold for the atomic to molecular gas transition (Krumholz et al. 2009; Sternberg et al. 2014).  High-redshift galaxies significantly exceed this threshold surface density, and are therefore expected to be entirely dominated by gas in the molecular phase within their visible region (see also Bauermeister et al. 2010).  Including a radially flat HI distribution at the threshold surface density for all of our galaxies would increase the baryonic mass within 10 kpc by $3 \times 10^9\ M_{\sun}$.  The baryonic mass would consequently rise by a factor of 1.06 in the median, and a maximum increase of $\sim 30\%$ for the least massive galaxies in our sample.  We conclude that contributions from atomic gas may have a minor, but not dominant impact on the assessment of the mass budget within the central few scale lengths of early disks.  Moreover, the effect would again be to increase the overall baryonic mass, as would also be the case if a substantial fraction of the molecular hydrogen is not traced by CO (or dust) (see Bolatto et al. 2013 and references therein).  Wolfire et al. (2010) estimate that for solar metallicity, and for the typical UV fields, densities and column densities in high-z SFGs, the mass fraction of this 'CO-dark' gas may be between 20 and 30\%.

In contrast to the above effects, a bias yielding over- rather than underestimated $M_{gas}$ would have to be invoked to explain the presence of $f_{bar} > 1$ galaxies.  The gas scaling relations we adopted were parameterized in terms of redshift, stellar mass and star formation rate, without further regard to galaxy size.  For the dust-based method it relied on stacked far-infrared SEDs in bins of (z, SFR, $M_{star}$).  While galaxy size itself shows systematic variations depending on these three parameters (Wuyts et al. 2011b), significant scatter in size remains within each (z, SFR, $M_{star}$) bin.  The most compact and therefore highest surface density ones within each bin necessarily have the shortest dynamical time, and may therefore be expected to have a higher star formation efficiency than average (Genzel et al. 2010).  Likewise, they may feature a higher dust temperature than the average {\it Herschel} stacked SED, and potentially a higher excitation and/or lower conversion factor in the case of the CO-based gas method.  A higher star formation efficiency for such dense systems would translate to lower gas reservoirs and hence a reduced tension with dynamical constraints.

%%%%%
% FIG 10
%%%%%
\begin {figure}[t]
\plotone{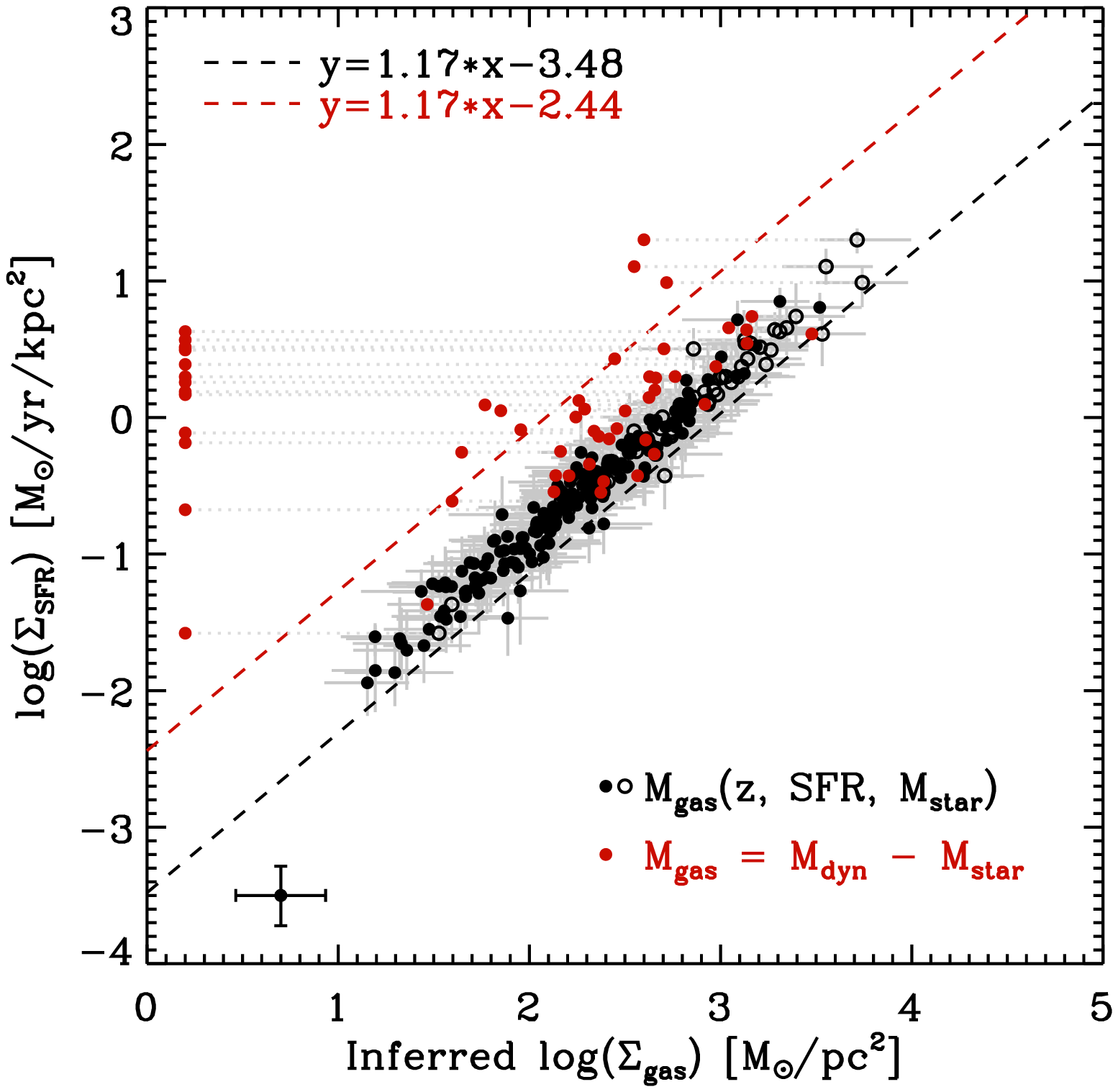}
\caption{Kennicutt-Schmidt relation of SFGs in our KMOS$^{\rm 3D}$ sample.  Black symbols indicate the location of SFGs in the KS diagram when gas masses are inferred from the galaxies' redshift, stellar mass and SFR, following our default CO- and dust-based gas scaling relations, with empty black circles marking objects for which $M_{bar} > M_{dyn}$.  Red circles represent the location of the latter subset if adopting the difference between dynamical and stellar mass as estimate of the gas content instead.  The majority of these predominantly high surface density systems then occupy star formation efficiencies ranging from the 'normal SFG' ({\it black dashed line}) to ULIRG/SMG ({\it red dashed line}) regime, as identified by Genzel et al. (2010).  Cases for which $M_{star} > M_{dyn}$ are positioned on the far left of the diagram.
\label{KS.fig}}
\vspace{-4mm}
\end{figure}

In Figure\ \ref{KS.fig}, we briefly explore what star formation efficiencies would have to be invoked in order to bring the baryonic mass budget into agreement with the dynamical constraints.  That is, we contrast the location of $f_{bar} > 1$ galaxies on the Kennicutt-Schmidt (KS) diagram for gas reservoirs inferred from the CO- and dust-based scaling relations (empty black circles), to an alternative realization where we neglect for simplicity the presence of dark matter, and equate $M_{gas} = M_{dyn} - M_{star}$ (red filled circles).  This exercise is akin to Downes \& Solomon (1998) who used dynamical mass estimates to constrain gas masses and star formation efficiencies in nearby Ultra-Luminous Infra-Red Galaxies.  Naturally, any presence of dark matter within the inner regions of the galaxy would boost the star formation efficiency implied by the latter method.

With the exception of a dozen outliers which already have $M_{star} > M_{dyn}$, the majority of $f_{bar} > 1$ objects now range from the KS relation for 'normal SFGs' defined by Genzel et al. (2010; dashed black line) to the KS relation for the 'ULIRG/SMG regime' (dashed red line).  This thought experiment encourages the exploration of size dependence in future studies of gas scaling relations through cosmic time, especially now that far infrared sizes (rather than the $H$-band sizes adopted here) are within reach with the high-resolution capabilities of ALMA and PdBI/NOEMA.

Figure\ \ref{zevol_parent.fig} (top-right panel) painted an evolutionary picture in which the same population followed over cosmic time featured a relatively constant $f_{star}$.  We point out that the observed increase of $f_{bar}$ with redshift (Figure\ \ref{zevol_parent.fig}, bottom-right panel) is therefore inherently linked to the larger gas reservoirs in disk galaxies at early times, inferred from our scaling relation methodology to derive $M_{gas}$.  On the one hand, it is encouraging that independent observations of CO and dust continuum tracers, the combination of which is parameterized by the adopted scaling relation, consistently suggest rapidly rising gas fractions with lookback time, as also anticipated from enhanced cosmological accretion rates at high redshift (see, e.g., Dekel et al. 2009).  On the other hand, we conclude that the presence of objects with $f_{bar} > 1$ may reveal limitations of this parameterization.

\subsection{Comparison to Illustris simulation}
\label{discussion_illustris.sec}

We now place our observational census of the mass budget in early disks in context, by contrasting our findings to expectations from a state-of-the-art simulation of galaxy formation in a $\Lambda$CDM cosmology, and to measurements of baryonic fractions within galaxies in the local universe.

To this end, we make use of the public data release of the Illustris Simulation (Vogelsberger 2014a,b; Nelson, D. et al. 2015).  Illustris is a large volume (106.5 Mpc$^3$) cosmological hydrodynamical simulation run with the moving-mesh code Arepo (Springel 2010).  Through a set of physical models it follows self-consistently the evolution of galaxies from z = 127 to the present day (Genel et al. 2014), and the interplay between their dark matter, gas, and stellar components.  We refer the reader to Genel et al. (2015) and Pillepich et al. (2014) for a more in depth discussion of the relation between baryons and the dark matter halos that host them in Illustris, and to Schaller et al. (2015) and Zavala et al. (2016) for a discussion on the topic based on an alternative cosmological hydrodynamical simulation EAGLE (Schaye et al. 2015; Crain et al. 2015).

%%%%%
% FIG 11
%%%%%
\begin {figure}[t]
\plotone{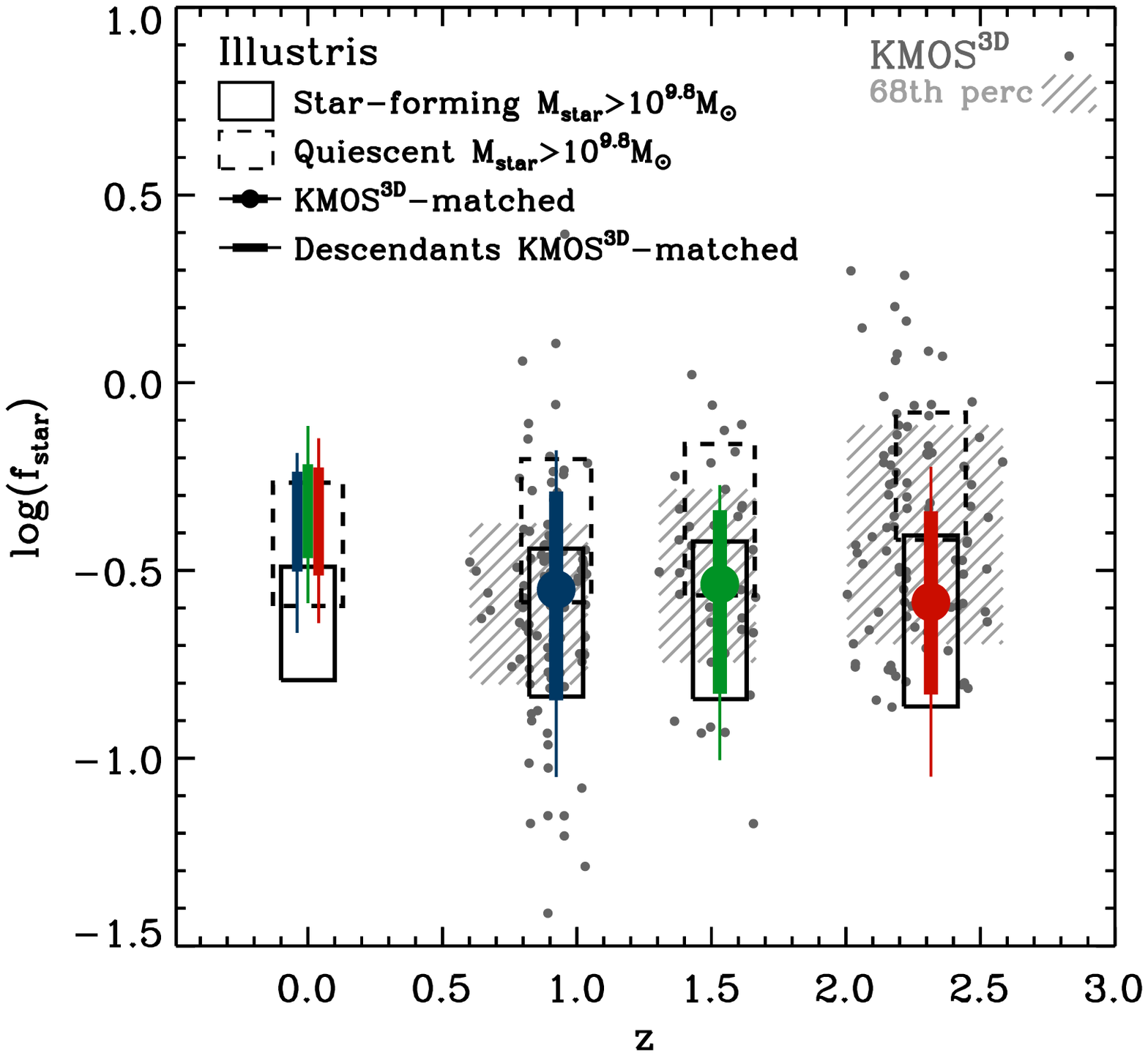}
\plotone{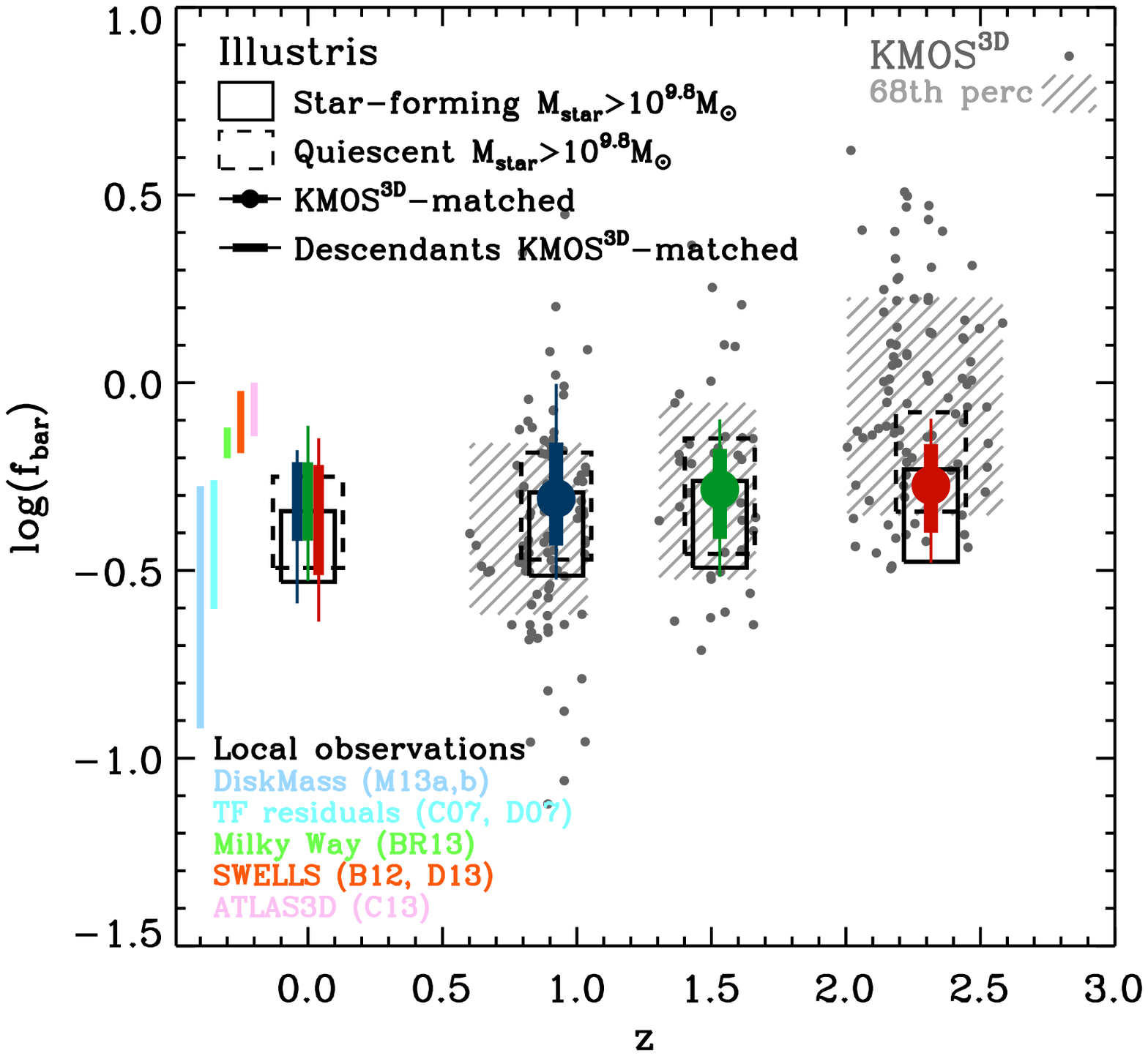}
\caption{
{\it Top:} Stellar mass fractions within the stellar half-mass radius as a function of redshift as simulated in Illustris.  Boxes indicate the central 68th percentiles for star-forming ($SSFR > 0.7/t_{\rm Hubble}$; {\it solid}) and quiescent ($SSFR < 0.7/t_{\rm Hubble}$; {\it dashed}) galaxies.  Central 68th and 90th percentiles for a simulated galaxy sample matched to the star formation rate and stellar mass distribution of our KMOS$^{\rm 3D}$ sample are marked in color.  Also shown are their descendant population at z = 0.  KMOS$^{\rm 3D}$ observations are overplotted with grey dots for reference.  {\it Bottom:} Idem, but showing the baryonic mass fractions within the stellar half-mass radius.
\label{Illustris_z.fig}}
\vspace{-0.3cm}
\end{figure}

%%%%%
% FIG 12
%%%%%
\begin {figure}[htbp]
\plotone{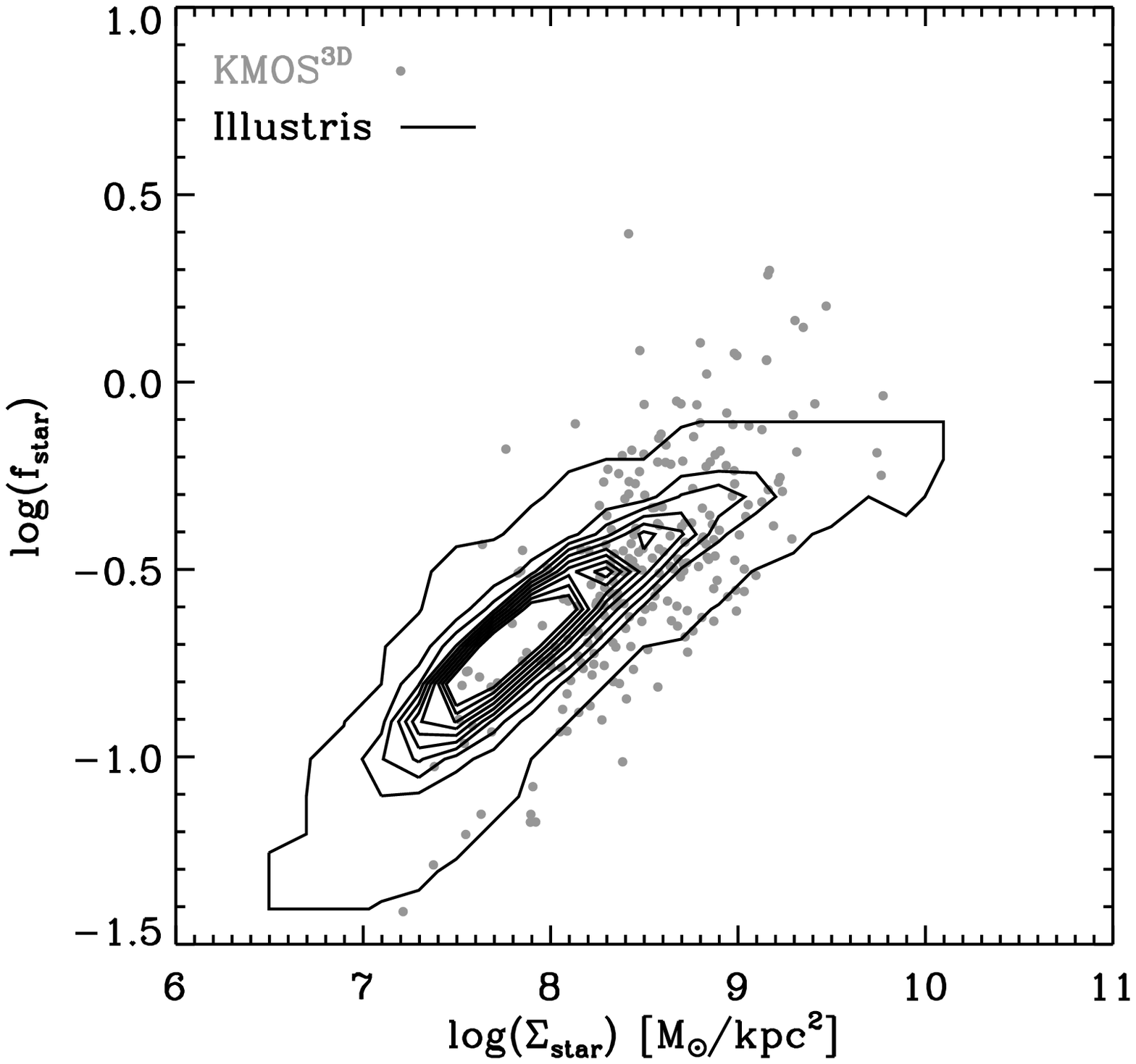}
\plotone{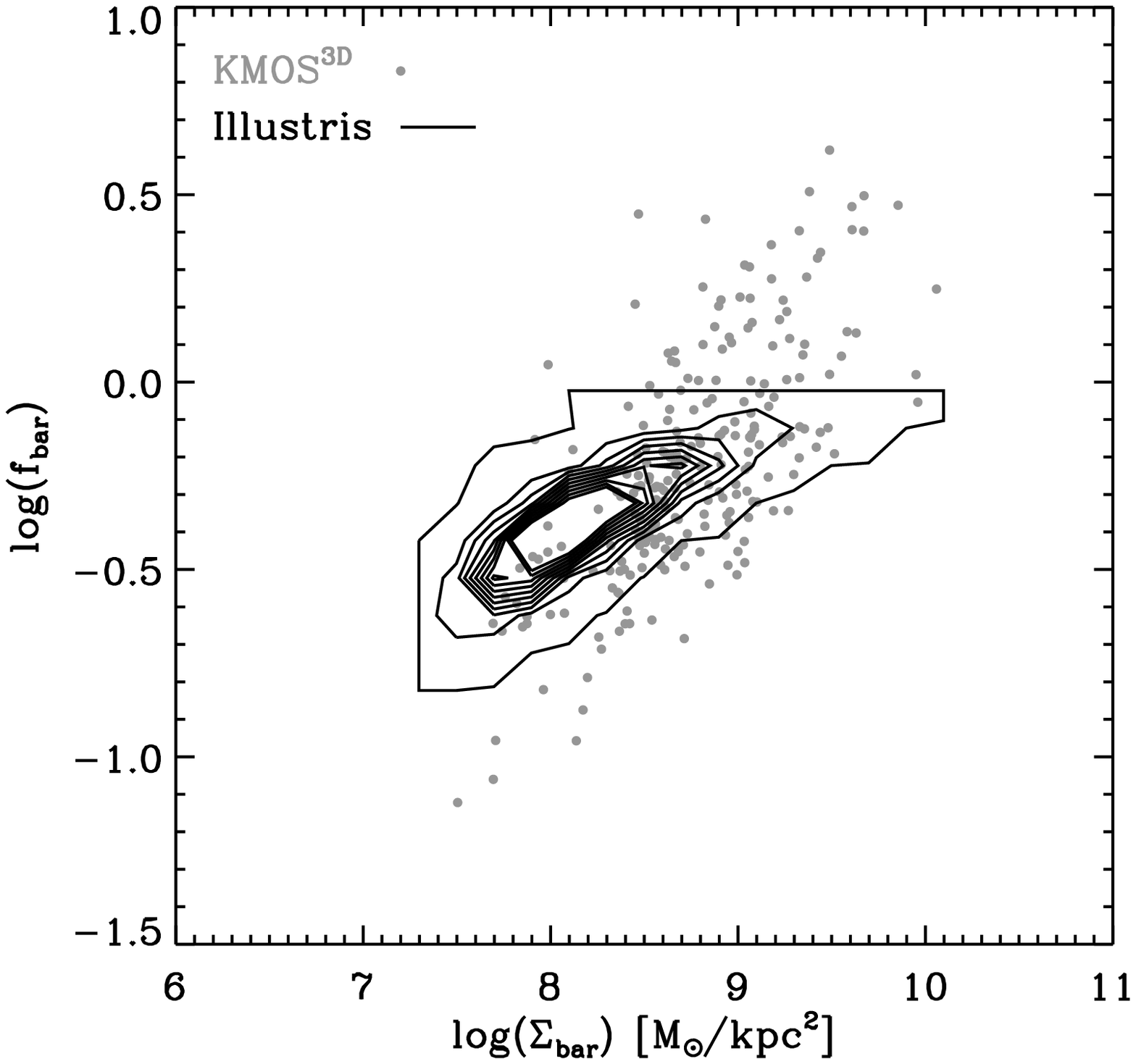}
\caption{
Stellar ({\it Left}) and baryonic ({\it right}) mass fractions within the stellar half-mass radius as a function of stellar and baryonic surface mass density, as simulated in Illustris.  KMOS$^{\rm 3D}$ observations are overplotted with grey dots for reference.  Both observations and simulations show a strong relation between mass fractions and surface densities.
\label{Illustris_surfdens.fig}}
\vspace{-0.3cm}
\end{figure}

%%%%%
% FIG 13
%%%%%
\begin {figure}[htbp]
\plotone{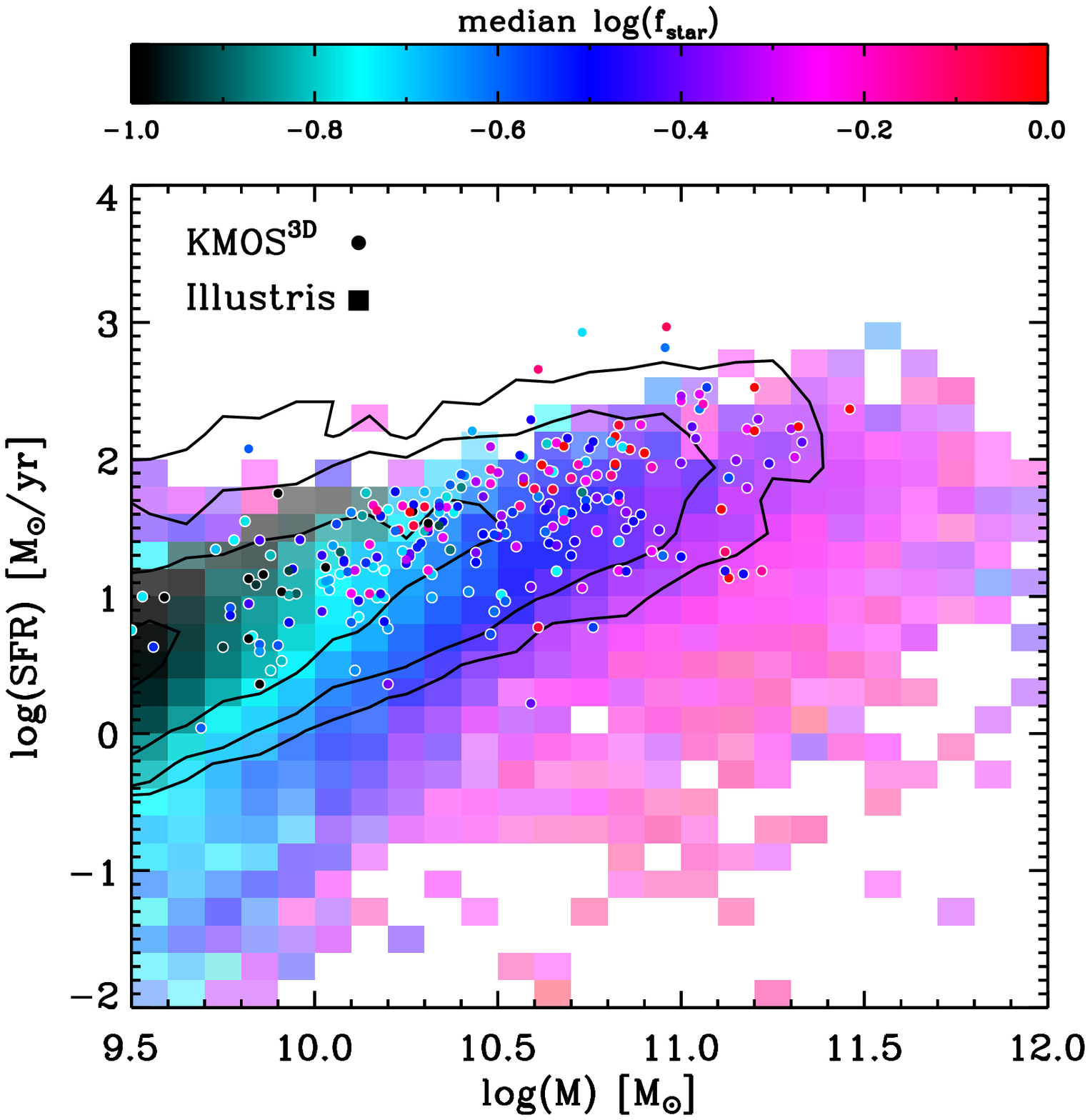}
\plotone{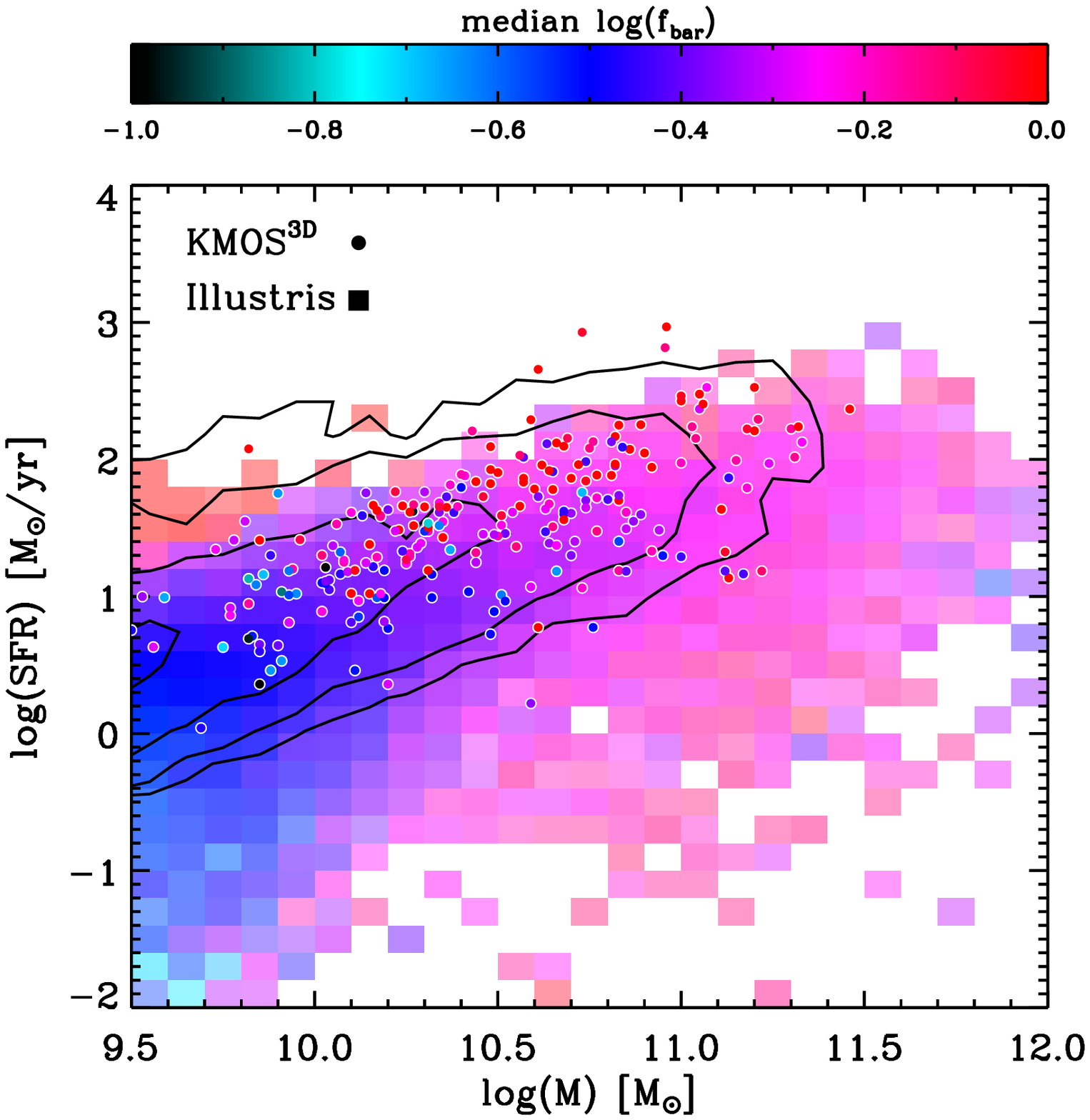}
\caption{
{\it Top:} Stellar mass fractions in the SFR-Mass diagram.  Color bins represent the full Illustris galaxy population (star-forming and quiescent) extracted from snapshots corresponding to redshifts z = 0.9, 1.5, and 2.3.  Systematic variations in the median $f_{star}(<R_e)$ of simulated galaxies are present across the diagram.  Qualitatively similar trends are notable for the observed KMOS$^{\rm 3D}$ sample at $0.6 < z < 2.6$.  Black lines mark iso-density contours of the underlying 3D-HST SFG population.  {\it Bottom:} Idem for baryonic mass fractions, where we capped the color scheme at the physical limit $log(f_{bar})=0$.
\label{Illustris_SFRM.fig}}
\vspace{-0.5cm}
\end{figure}

Figure\ \ref{Illustris_z.fig} illustrates the distribution of stellar and baryonic mass fractions within $R_e$ (defined as the stellar half-mass radius) for snapshots of the Illustris simulation corresponding to redshifts 0, 0.9, 1.5, and 2.3.  Considering the simulated galaxy population more massive than $10^{9.8}\ M_{\sun}$, no significant dependence of the stellar and baryonic mass fractions on redshift is found for SFGs (defined as $SSFR > 0.7/t_{\rm Hubble}$).  At all redshifts, the mass fractions in quiescent galaxies, and particularly the contribution of the stellar component, is elevated with respect to that of the star-forming population.  Since our KMOS$^{\rm 3D}$ sample does not represent a random drawing from the galaxy stellar mass function, but features SFGs with a relatively flat mass distribution (see Section\ \ref{sample.sec}), we carry out a more consistent comparison by creating a ($SFR,\ M_{star}$)-matched sample containing 10 Illustris galaxies for each observed KMOS$^{\rm 3D}$ galaxy.  Their distribution of stellar and baryonic mass fractions is illustrated in blue, green and red colors, for the three high redshift bins.  At $z \sim 0.9$ and $z \sim 1.5$, we find $f_{star}$ and $f_{bar}$ distributions that are consistent with our observations.  At $z \sim 2.3$, the KMOS$^{\rm 3D}$-matched Illustris sample only overlaps with the lower tail of the observed distribution, hence representing a clear deviation from the observed trend.  If we were to use galaxy size as an additional parameter (together with $SFR$ and $M_{star}$) in constructing a KMOS$^{\rm 3D}$-matched Illustris sample, no noticeable difference is found at $z \sim 2.3$, whereas simulated mass fractions for the matched sample at $z \sim 0.9$ and $z \sim 1.5$ increase by $\sim 0.1$ dex in the median, with a modest reduction in scatter.  This reflects the fact that at these redshifts the distribution of SFG sizes in Illustris extends to larger systems than observed in the real universe.

Using the SubLink merger trees provided in the Illustris public data release (Rodriguez-Gomez et al. 2015), we trace the z = 0 descendants of the KMOS$^{\rm 3D}$-matched sample and find the breakdown of their mass budget within $R_e$ to be more akin to that of the quiescent z = 0 population than the star-forming one\footnote{No appreciable change in descendant mass fractions is noted when adopting galaxy size as an additional matching parameter.}.  In this light, it is also interesting to look at observational measurements of the baryonic mass fraction within nearby galaxies, as compiled by Courteau \& Dutton (2015)\footnote{Courteau \& Dutton (2015) evaluate the mass fraction at 2.2 $R_d \sim 1.3 R_e$.}.  These include disk galaxies of the DiskMass survey (Martinsson et al. 2013a,b; M13a,b), studies of Tully-Fisher residuals by Courteau et al. (2007; C07) and Dutton et al. (2007; D07), an analysis of Milky Way kinematics by Bovy \& Rix (2013; BR13), the SWELLS sample of massive gravitationally lensed spirals (Barnab\`{e} et al. 2012, B12; Dutton et al. 2013, D13), and the ATLAS$^{\rm 3D}$ early-type galaxy sample (Cappellari et al. 2013; C13).  While the baryonic mass fractions observed at $z > 2$ in KMOS$^{\rm 3D}$ tend to exceed those of intermediate mass nearby disks (M13a,b; C07; D07), they are in the range of what is observed in the most massive nearby spirals (SWELLS) and the local early-type galaxy population (ATLAS$^{\rm 3D}$).  The latter are the more likely descendants of our high-z SFGs, according to the Illustris simulation, but also based on simpler co-moving number density arguments (van Dokkum et al. 2010; Muzzin et al. 2013).

In Figure\ \ref{Illustris_surfdens.fig}, we contrast one of our main findings, namely that the mass fractions correlate strongly with measures of surface density, to the equivalent relations for the Illustris simulated galaxy population over the same redshift range.  Clearly, correlations of $f_{star}$ with $\Sigma_{star}$ and $f_{bar}$ with $\Sigma_{bar}$ as discussed for our observations in Section\ \ref{surfdens.sec} are also inherently present in hydrodynamical simulations that model the assembly of gas and build up of stars within dark matter halos formed according to a $\Lambda$CDM cosmology.  Quantitatively, however, significant differences are notable.  The equivalent relations to those described by equations\ \ref{eqn_1} -\ \ref{eqn_3} for the observed galaxy population feature shallower slopes for the simulated galaxies, of $f_{star} \sim \Sigma_{star}^{0.41}$, $f_{bar} \sim \Sigma_{gas}^{0.31}$, and $f_{bar} \sim \Sigma_{bar}^{0.27}$.  These correspond to differences at the $\sim 2\sigma$ and $> 3\sigma$ level for the stellar and baryonic mass fraction relations, respectively, and may be related to differences in the galaxy size distribution discussed by Snyder et al. (2015).  At low surface densities, the simulated galaxies typically feature higher mass fractions than observed, whereas at higher surface densities systems with mass fractions in the unphysical ($> 1$) regime are trivially absent.

Finally, we consider in Figure\ \ref{Illustris_SFRM.fig} how the observed and simulated galaxies vary in $f_{star}$ and $f_{bar}$ depending on their position in the SFR - stellar mass plane.  As is the case for many other galaxy properties related to structure, mode of star formation, gas and dust properties (e.g., Wuyts et al. 2011b; Magnelli et al. 2014; Genzel et al. 2015), systematic variations are notable.  In part, these reflect the above-described correlations with surface density.  For example, the iso-$f_{star}$ contours of simulated galaxies in the top panel of Figure\ \ref{Illustris_SFRM.fig} coincide more or less with lines of constant stellar surface mass density (see also Brennan et al. 2016 for a comparison of observed $\Sigma_{star}$ in the SFR-Mass plane versus the equivalent behavior in semi-analytic models of galaxy formation).

We conclude that overall, galaxies as simulated in state-of-the-art cosmological hydrodynamical simulations share many qualitative similarities with their counterparts in the real universe, in terms of their mass budget breakdown and relation to other galaxy parameters.  However, additional work on the interface between observations and simulations, ideally using cosmological boxes realized with a range of physics implementations, is desired to pin down the origin of factor of $\sim 2 - 3$ deviations in certain areas of parameter space (e.g., the high observed $f_{bar}$ at $z > 2$).

\section{Summary}
\label{summary.sec}

Multi-object spectrographs such as the 24-IFU KMOS instrument on the VLT are opening a window on the dynamical mass budget of distant galaxies for samples of increasing statistical significance.  In this paper, we carried out detailed dynamical modeling of 240 massive ($\log(M_{star}) \gtrsim 9.8$) star-forming disks from the KMOS$^{\rm 3D}$ survey, spanning a wide redshift range of $0.6 < z < 2.6$ and extending in the median out to $\sim 9.5$ kpc.  Our main conclusions are the following:

$\bullet$ Over the full redshift range, distant star-forming disk galaxies leave significant room for other mass components than stars within their visible extent.  Adopting a Chabrier (2003) IMF, stars on average account for a third of the total mass budget.

$\bullet$ Folding in molecular gas masses derived from CO- and dust-based gas scaling relations, we find baryons (i.e., gas plus stars) to account for typically 56\% of the total mass budget, increasing with redshift such that star-forming disk galaxies at $z > 2$ are fully baryon-dominated, with little room for significant dark matter contributions within their inner regions ($f_{bar} \sim 0.9$).

$\bullet$ In order to estimate the evolution for the overall underlying population, we compose a mass-complete sample of SFGs from the 3D-HST/CANDELS data set, and assign stellar and baryonic mass fractions to each based on the KMOS$^{\rm 3D}$ galaxy from our sample that is best matched in its intrinsic properties.  This approach does not significantly alter our conclusions.  Likewise, if adopting an evolving mass limit to trace the progenitors of today's $\log(M_{star,\ z \sim 0}) > 10.7$ galaxy population, a similar redshift evolution of baryonic mass fractions is inferred.

$\bullet$ The inference of missing mass components (particularly if only the stellar mass is contrasted to the dynamical constraints) cannot be attributed to systematics in the galaxy inclinations, estimated from axial ratio measurements.  An implausibly large number of edge-on systems, inconsistent with axial ratios and statistical expectations from random viewing angles, would have to be invoked to reproduce the observed range in rotational velocities.

$\bullet$ A large ($\sim 0.3$ dex) galaxy-to-galaxy scatter is noted in the stellar and baryonic mass fractions.  These variations are not random, but correlate with galaxy properties, most strongly so with measures of surface density.  Systems with compact stellar distributions feature the largest stellar mass fractions.  The highest baryonic mass fractions occur for galaxies with the highest inferred gas (and baryonic) surface densities.  The observed trends are in line with a scenario in which our kinematic tracer probes further into the dark matter halo in which the galaxy is embedded if the galaxy's stellar and baryonic distributions are extended, whereas in the case of compact and high surface density systems it is the heavily baryon-dominated inner region that is probed.  Similarly strong correlations between mass fractions within $R_e$ and surface density also follow naturally from the physical models with which baryons are traced in a $\Lambda$CDM context within the cosmological hydrodynamical simulation Illustris.  The presence of a significant population of observed disks which formally lie in the unphysical ($f_{bar} > 1$) regime suggests that in addition, an enhanced star formation efficiency may apply to more compact SFGs (see also Spilker et al. 2016).

Looking ahead, systematic observing campaigns of dust continuum and molecular line tracers with PdBI/NOEMA and ALMA sampling galaxies over a wide range of redshift, stellar population and structural properties will tighten the constraints on gas and hence baryonic mass estimates further, including also a direct assessment of its spatial distribution (e.g., Barro et al. 2016; Tadaki et al. in prep).  The exploration of outer rotation curves has the potential of providing a secondary, purely kinematic path towards tightening constraints on baryonic mass fractions in high-redshift disk galaxies (Lang et al. in prep; Genzel et al. in prep).

%\acknowledgments
\vspace{0.4cm}
We thank the staff at Paranal Observatory for their excellent support during numerous observing runs.  MF and DW acknowledge the support of the Deutsche Forschungs Gemeinschaft (DFG) via Project WI 3871/1-1.

\onecolumngrid %not needed in ApJ submission format

\vspace{0.2in}
\begin{center} APPENDIX A  TESTING THE RECOVERY OF DYNAMICAL MASSES\end{center}

\begin {figure*}[h]
\epsscale{0.92} %{1.0} in ApJ submission
\plotone{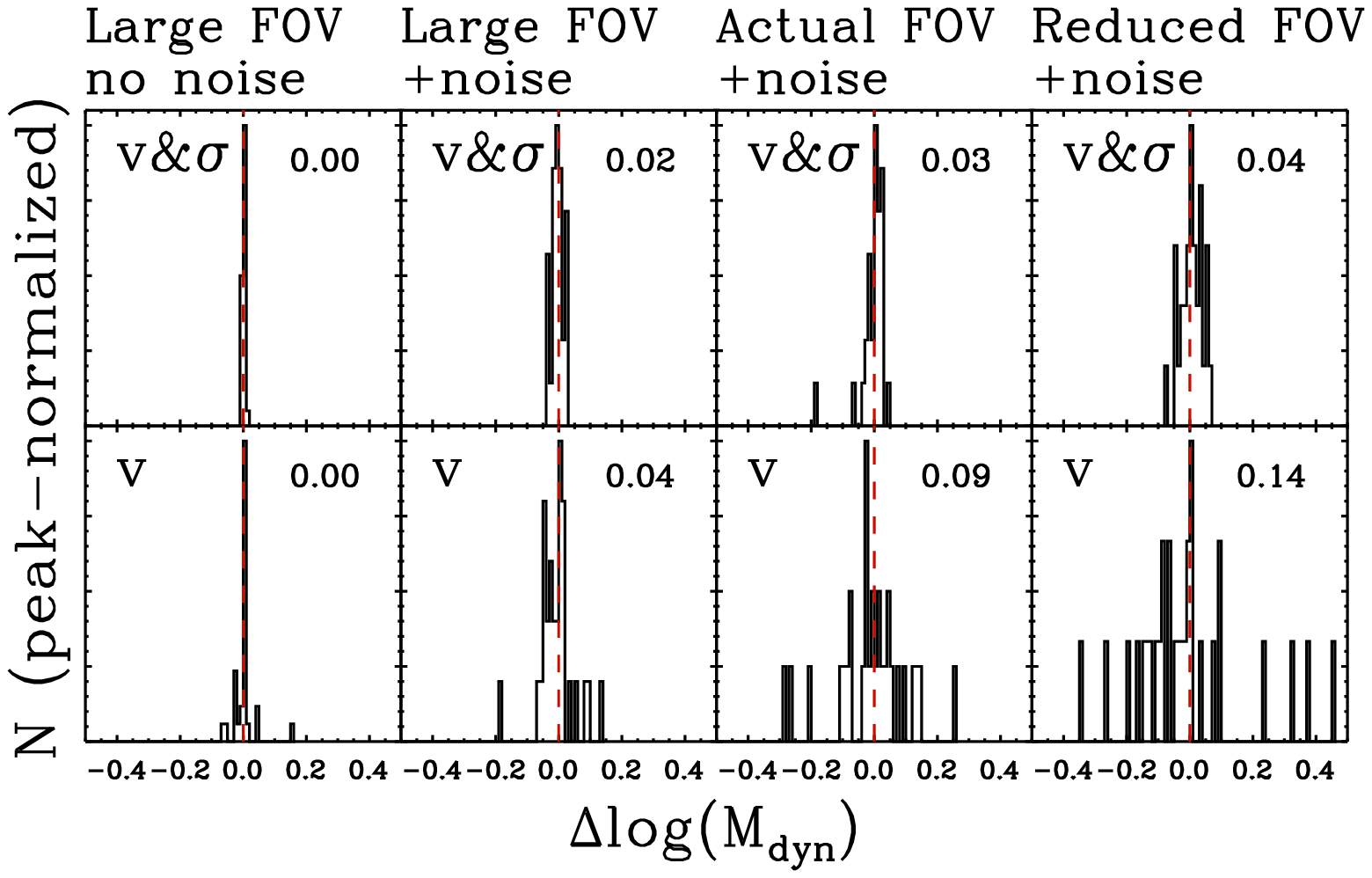}
\epsscale{1}
\caption{
Distribution of recovered minus intrinsic dynamical masses for mock disks with inclination, mass and size properties matched to those of the 31 largest galaxies in our sample.  Each panel contains the same number of objects, but histograms are normalized to the same peak height for more convenient presentation.  The normalized median absolution deviation of the distribution is listed in the top-right corner.  From left to right, panels differ in the radial extent of the kinematic profiles used in the fitting, and whether or not noise was applied to the mock disks.  The top row shows results for our default methodology, fitting velocity and dispersion profiles simultaneously.  The bottom row illustrates the effect of ignoring the dispersion information, leading to a broader distribution and hence poorer recovery of $M_{dyn}$.
\label{recover_Mdyn.fig}}
\end{figure*}

For some of the larger galaxies in our sample, the finite spatial coverage of our KMOS observations makes the radial extent of extracted kinematic profiles field-of-view limited rather than signal-to-noise limited.  We tested the potential impact on the recovered dynamical masses by taking the best-fit disk models for the 31 largest galaxies (those with $R_{e,H} > 0.8''$), applying noise at a level appropriate for our observations, and fitting them over the full radial extent, over the actual radial extent probed in our observations, and, for illustrative purposes, over a more severely restricted radial extent with the outer kinematic extractions at $r = 0.8''$.  Figure\ \ref{recover_Mdyn.fig} shows how the recovered dynamical mass compares to the intrinsic one known for these mock disks.  The top row shows that an encouraging match is obtained when following our methodology of fitting velocity and dispersion profiles simultaneously.  The bottom panels illustrate how the accuracy of the recovered $M_{dyn}$ degrades when no dispersion information is taken into account, especially if the rotation curve is not traced far out.

%%%%%
% FIG 14 [testing the recovery of Mdyn]
%%%%%

$ $\\
% References
\begin{references}
{\footnotesize

\reference{} Barnab\`{e}, M., Dutton, A. A., Marhsall, P. J., et al. 2012, MNRAS, 423, 1073
\reference{} Barnab\`{e}, M., Czoske, O., Koopmans, L. V. E., Treu, T.,\& Bolton, A. S. 2011, MNRAS, 415, 2215
\reference{} Barro, G., Kriek, M., P\'erez-Gonz\'alez, P. G., et al. 2016, ApJL, in press (arXiv1607.01011)
\reference{} Barro, G., Trump, J. R., Koo, D. C., et al. 2014, ApJ, 795, 145
\reference{} Bauermeister, A., Blitz, L.,\& Ma, C.-P. 2010, ApJ, 717, 323
\reference{} Bershady, M. A., Martinsson, T. P. K., Verheijen, M. A. W., et al. 2011, ApJ, 739, L47
\reference{} Berta, S., Lutz, D., Nordon, R., et al. 2013, A\&A, 555, 8
\reference{} B\'{e}thermin, M., Daddi, E., Magdis, G., et al. 2015, A\&A, 573, 113
\reference{} Bezanson, R., van Dokkum, P. G., van de Sande, J., Franx, M., Leja, J.,\& Kriek, M. 2013, ApJ, 779, 21
\reference{} Bigiel, F.,\& Blitz, L. 2012, ApJ, 756, 183
\reference{} Bolatto, A. D., Wolfire, M.,\& Leroy, A. K. 2013, ARA\&A, 51, 207
\reference{} Bosma, A. 1978, Ph.D. Thesis, Groningen Univ.
\reference{} Bovy, J.,\& Rix, H.-W. 2013, ApJ, 779, 115
\reference{} Bouch\'{e}, N., Carfantan, H., Schroetter, I., Michel-Dansac, L.,\& Contini, T. 2015, AJ, 150, 92
\reference{} Brammer, G. B., van Dokkum, P. G., Franx, M., et al. 2012, ApJS, 200, 13
\reference{} Brewer, B. J., Dutton, A. A., Treu, T., et al. 2012, MNRAS, 422, 3574
\reference{} Bruzual, G,\& Charlot, S. 2003, MNRAS, 344, 1000
\reference{} Burkert, A., F\"{o}rster Schreiber, N. M., Genzel, R., et al. 2016, ApJ, in press (arXiv1510.03262)
\reference{} Burkert, A., Genzel, R., Bouch\'{e}, N., et al. 2010, ApJ, 725, 2324
\reference{} Capozzi, D., Maraston, C., Daddi, E., Renzini, A., Strazzullo, V.,\& Gobat, R. 2016, MNRAS, 456, 790
\reference{} Cappellari, M. 2016, ARA\&A, 54 (arXiv1602.04267)
\reference{} Cappellari, M., Scott, N., Alatalo, K., et al. 2013, MNRAS, 432, 1709
\reference{} Cappellari, M., McDermid, R. M., Alatalo, K., et al. 2012, Nature, 484, 485
\reference{} Carilli, C. L.,\& Walter, F. 2013, ARA\&A, 51, 105
\reference{} Chabrier, G. 2003, PASP, 115, 763
\reference{} Cole, S., Norberg, P., Baugh, C. M., et al. 2001, MNRAS, 326, 255
\reference{} Conroy, C.,\& van Dokkum, P. G., 2012, ApJ, 760, 71
\reference{} Contini, T., Epinat, B., Bouch\'e, N. et al. 2016, A\&A, 591, 49
\reference{} Courteau, S.,\& Dutton, A. A. 2015, ApJ, 801, 20
\reference{} Courteau, S., Dutton, A. A., van den Bosch, F. C., et al. 2007, ApJ, 671, 203 %Ill
\reference{} Crain, R. A., Schaye, J., Bower, R. G., et al. 2015, MNRAS, 450, 1937 %Ill
\reference{} Cresci, G., Hicks, E. K. S., Genzel, R., et al. 2009, ApJ, 697, 115
\reference{} Daddi, E., Elbaz, D., Walter, F., et al. 2010b, ApJ, 714, 118
\reference{} Daddi, E., Bournaud, F., Walter, F., et al. 2010a, ApJ, 713, 686
\reference{} Daddi, E., Dickinson, M., Morrison, G., et al. 2007, ApJ, 670, 156
\reference{} Davies, R. I., Agudo Berbel, A., Wiezorrek, E., et al. 2013, A\&A, 558, 56
\reference{} Davies, R. I., F\"{o}rster Schreiber, N. M., Cresci, G., et al. 2011, ApJ, 741, 69
\reference{} Davies, R. I., Maciejewski, W., Hicks, E. K. S., Tacconi, L. J., Genzel, R.,\& Engel, H. 2009, ApJ, 702, 114
\reference{} Downes, D.,\& Solomon, P. M. 1998, ApJ, 507, 615
\reference{} Dutton, A. A., Treu, T., Brewer, B. J., et al. 2013, MNRAS, 428, 3183 %Ill
\reference{} Elbaz, D., Daddi, E., Le Borgne, D., et al. 2007, A\&A, 468, 33
\reference{} Ferreras, I., La Barbera, F., de la Rosa, I. G., et al. 2013, MNRAS, 429, 15
\reference{} F\"{o}rster Schreiber, N. M., Shapley, A. E., Erb, D. K., Genzel, R., Steidel, C. C., Bouch\'{e}, N., Cresci, G.,\& Davies, R. 2011, ApJ, 731, 65
\reference{} F\"{o}rster Schreiber, N. M., Genzel, R., Bouch\'{e}, N., et al. 2009, ApJ, 706, 1364
\reference{} Freundlich, J., Combes, F., Tacconi, L. J., et al. 2013, A\&A, 553, 130
\reference{} Genel, S., Fall, S. M., Hernquist, L., et al. 2015, ApJ, 804, 40 %Ill
\reference{} Genel, S., Vogelsberger, M., Springel, V., et al. 2014, MNRAS, 445, 175 %Ill
\reference{} Genzel, R., Tacconi, L. J., Lutz, D., et al. 2015, ApJ, 800, 20
\reference{} Genzel, R., F\"{o}rster Schreiber, N. M., Lang, P., et al. 2014, ApJ, 785, 75
\reference{} Genzel, R., Tacconi, L. J., Kurk, J., et al. 2013, ApJ, 773, 68
\reference{} Genzel, R., Tacconi, L. J., Combes, F., et al. 2012, ApJ, 746, 69
\reference{} Genzel, R., Tacconi, L. J., Gracia-Carpio, J., et al. 2010, MNRAS, 407, 2091
\reference{} Glazebrook, K. 2013, PASA, 30, 56
\reference{} Grogin, N. A., Kocevski, D. D., Faber, S. M., et al. 2011, ApJS, 197, 35
\reference{} Groves, B. A., Schinnerer, E., Leroy, A., et al. 2015, ApJ, 799, 96
\reference{} Ilbert, O., McCracken, H. J., Le F\`{e}vre, O., et al. 2013, A\&A, 556, 55
\reference{} Kartaltepe, J. S., Mozena, M., Kocevski, D., et al. 2015, ApJS, 221, 11
\reference{} Kassin, S. A., Weiner, B. J., Faber, S. M., et al. 2012, ApJ, 758, 106
\reference{} Kennicutt, R. C., Jr. 1998, ApJ, 498, 541
\reference{} Keres, D., Yun, M. S.,\& Young, J. S. 2003, ApJ, 582, 659
\reference{} Koekemoer, A. M., Faber, S. M., Ferguson, H. C., 2011, ApJS, 197, 36
\reference{} Kriek, M., Labb\'e, I., Conroy, C., et al. 2010, ApJ, 722, 64
\reference{} Krumholz, M. R., Leroy, A. K.,\& McKee, C. F. 2011, ApJ, 731, 25
\reference{} Krumholz, M. R., McKee, C. F.,\& Tumlinson, J. 2009, ApJ, 693, 216
\reference{} Lang, P., Wuyts, S., Somerville, R. S., et al. 2014, ApJ, 788, 11
\reference{} Leroy, A. K., Bolatto, A., Gordon, K., et al. 2011, ApJ, 737, 12
\reference{} Leroy, A. K., Walter, F., Bigiel, F., et al. 2009, AJ, 137, 4670
\reference{} Lutz, D., Poglitsch, A., Altieri, B., et al. 2011, A\&A, 532, 90
\reference{} Magnelli, B., Lutz, D., Saintonge, A., et al. 2014, A\&A, 561, 86 %Ill
\reference{} Magnelli, B., Popesso, P., Berta, S., et al. 2013, A\&A, 553, 132
\reference{} Magnelli, B., Saintonge, A., Lutz, D., et al. 2012, A\&A, 548, 22
\reference{} Mandelker, N., Dekel, A., Ceverino, D., Tweed, D., Moody, C. E.,\& Primack, J. 2014, MNRAS, 433, 3675
\reference{} Martin, A. M., Papastergis, E., Giovanelli, R., Haynes, M. P., Springob, C. M.,\& Stierwalt, S. 2010, ApJ, 723, 1359
\reference{} Martinsson, T. P. K., Verheijen, M. A. W., Westfall, K. B., Bershady, M. A., Andersen, D. R.,\& Swaters, R. A. 2013a, A\&A, 557, 130
\reference{} Martinsson, T. P. K., Verheijen, M. A. W., Westfall, K. B., Bershady, M. A., Andersen, D. R.,\& Swaters, R. A. 2013b, A\&A, 557, 131
\reference{} Maraston, C. 2005, MNRAS, 362, 799
\reference{} McCracken, H. J., Milvang-Jensen, B., Dunlop, J., et al. 2012, A\&A, 544, 156
\reference{} Mo, H. J., Mao, S.,\& White, S. D. M. 1998, MNRAS, 295, 319
\reference{} Momcheva, I. G., Brammer, G. B., van Dokkum, P. G., et al. 2016, ApJS, in press (arXiv1510.02106)
\reference{} Muzzin, A., Marchesini, D., Stefanon, M., et al. 2013, ApJ, 777, 18
\reference{} Narayanan, D.,\& Krumholz, M. 2014, MNRAS, 442, 1411
\reference{} Nelson, D., Pillepich, A., Genel, S., et al. 2015, A\&C, 13, 12 %Ill
\reference{} Nelson, E. J., van Dokkum, P. G., F\"{o}rster Schreiber, N. M., et al. 2016, ApJ, in press (arXiv1507.03999)
\reference{} Nelson, E. J., van Dokkum, P. G., Momcheva, I., et al. 2013, ApJ, 763, 16
\reference{} Newman, A. B., Belli, S.,\& Ellis, R. S. 2015, ApJL, 813, 7
\reference{} Noeske, K. G., Weiner, B. J., Faber, S. M., et al. 2007, ApJ, 660, 43
\reference{} Noordermeer, E. 2008, MNRAS, 385, 1359
\reference{} Papovich, C., Finkelstein, S. L., Ferguson, H. C., Lotz, J. M.,\& Giavalisco, M. 2011, MNRAS, 412, 1123
\reference{} Patel, S. G., van Dokkum, P. G., Franx, M., et al. 2013, ApJ, 766, 15
\reference{} Peng, C. Y., Ho, L. C., Impey, C. D.,\& Rix, H.-W. 2010, AJ, 139, 2097
\reference{} Pillepich, A., Vogelsberger, M., Deason, A., et al. 2014, MNRAS, 444, 237 %Ill
\reference{} Price, S. H., Kriek, M., Shapley, A. E., et al. 2016, ApJ, 819, 80
\reference{} R\'{e}my-Ruyer, A., et al. 2014, A\&A, 563, 31
\reference{} Rix, H.-W., Guhathakurta, P., Colless, M.,\& Ing, K. 1997, MNRAS, 285, 779
\reference{} Rodriguez-Gomez, V., Genel, S., Vogelsberger, M., et al. 2015, MNRAS, 449, 49
\reference{} Rubin, V. C., Thonnard, N.,\& Ford, W. K., Jr. 1978, ApJ, 225, L107
\reference{} Salpeter, E. E. 1955, ApJ, 121, 161
\reference{} Santini, P., Maiolino, R., Magnelli, B., et al. 2014, A\&A, 562, 30
\reference{} Schaller, M., Frenk, C. S., Bower, R. G., et al. 2015, MNRAS, 452, 343
\reference{} Schaye, J., Crain, R. A., Bower, R. G., et al. 2015, MNRAS, 446, 521
\reference{} Scoville, N., Aussel, H., Sheth, K., et al. 2014, ApJ, 783, 84
\reference{} Skelton, R. E., Whitaker, K. E., Momcheva, I. G., et al. 2014, ApJS, 214, 24
\reference{} Snyder, G. F., Torrey, P., Lotz, J. M., et al. 2015, MNRAS, 454, 1886
\reference{} Spilker, J. S., Bezanson, R., Marrone, D. P., Weiner, B. J., Whitaker, K. E.,\& Williams, C. C., ApJ, submitted (arXiv1607.01785)
\reference{} Spiniello, C., Trager, S. C., Koopmans, L. V. E.,\& Chen, Y. P. 2012, ApJ, 753, 32
\reference{} Springel, V. 2010, MNRAS, 401, 791 %Ill
\reference{} Sternberg, A., Le Petit, F., Roueff, E.,\& Le Bourlot, J. 2014, ApJ, 790, 10
\reference{} Stott, J. P., Swinbank, A. M., Johnson, H. L., et al. 2016, MNRAS, 457, 1888
\reference{} Swaters, R. A., Verheijen, M. A. W., Bershady, M. A.,\& Andersen, D. R. 2003, ApJ, 587, 19
\reference{} Tacconi, L. J., Neri, R., Genzel, R., et al. 2013, ApJ, 768, 74
\reference{} Tacconi, L. J., Genzel, R., Neri, R., et al. 2010, Nature, 463, 781
\reference{} Tacchella, S., Carollo, C. M., Renzini, A., et al. 2015, Science, 348, 314
\reference{} Thomas, J., Saglia, R. P., Bender, R., et al. 2011, MNRAS, 415, 545
\reference{} Tomczak, A. R., Quadri, R. F., Tran, K. H., et al. 2014, ApJ, 783, 85
\reference{} Torrey, P., Wellons, S., Machado, F., et al. 2015, MNRAS, 454, 2770
\reference{} van der Kruit, P. C.,\& Allen, R. J. 1978, ARA\&A, 16, 103
\reference{} van der Marel, R. P. 1991, MNRAS, 253, 710
\reference{} van der Wel, A., Chang, Y.-Y., Bell, E. F., et al. 2014, ApJ, 792, 6
\reference{} van der Wel, A., Bell, E. F., H\"{a}ussler, B., et al. 2012, ApJS, 203, 24
\reference{} van de Sande, J., Kriek, M., Franx, M., et al. 2013, ApJ, 771, 85
\reference{} van Dokkum, P. G., Nelson, E. J., Franx, M., et al. 2015, ApJ, 813, 23
\reference{} van Dokkum, P. G., Leja, J., Nelson, E. J., et al. 2013, ApJ, 771, 35
\reference{} van Dokkum, P. G.,\& Conroy, C. 2010, Nature, 468, 940
\reference{} van Dokkum, P. G., Whitaker, K. E., Brammer, G., et al. 2010, ApJ, 709, 1018 %Ill
\reference{} Vogelsberger, M., Genel, S., Springel, V., et al. 2014a, Nature, 509, 177 %Ill
\reference{} Vogelsberger, M., Genel, S., Springel, V., et al. 2014b, MNRAS, 444, 1518 %Ill
\reference{} Walter, F., Brinks, E., de Blok, W. J. G., Bigiel, F., Kennicutt, R. C., Jr., Thornley, M. D.,\& Leroy, A. 2008, AJ, 136, 2563
\reference{} Whitaker, K. E., Franx, M., Leja, J., et al. 2014, ApJ, 795, 104
\reference{} Wisnioski, E., F\"{o}rster Schreiber, N. M., Wuyts, S., et al. 2015, ApJ, 799, 209
\reference{} Wolfire, M. G., Hollenbach, D., McKee, C. F. 2010, ApJ, 716, 1191
\reference{} Wuyts, S., F\"{o}rster Schreiber, N. M., Genzel, R., et al. 2012, ApJ, 753, 114
\reference{} Wuyts, S., F\"{o}rster Schreiber, N. M., van der Wel, A., et al. 2011b, ApJ, 742, 96
\reference{} Wuyts, S., F\"{o}rster Schreiber, N. M., Lutz, D., et al. 2011a, ApJ, 738, 106
\reference{} Wuyts, S., Franx, M., Cox, T. J., et al. 2009, ApJ, 696, 348
\reference{} Zavala, J., Frenk, C. S., Bower, R., et al. 2016, MNRAS, 460, 4466 %Ill
\reference{} Zibetti, S., Gallazzi, A., Charlot, S., Pierini, D.,\& Pasquali, A. 2013, MNRAS, 428, 1479
\reference{} Zibetti, S., Charlot, S.,\& Rix, H.-W. 2009, MNRAS, 400, 1181

}
\end {references}

\end {document}